# Revealing the Anisotropic Thermal Conductivity of Black Phosphorus using the Time-Resolved Magneto-Optical Kerr Effect

*Jie Zhu,[1,5] Haechan Park,[3] Jun-Yang Chen,[2] Xiaokun Gu,[4] Hu Zhang,[1,6] Sreejith Karthikeyan,[2] Nathaniel Wendel,[2] Stephen A. Campbell,[2] Matthew Dawber,[3] Xu Du,[3] Mo Li,[2] Jian-Ping Wang,[2] Ronggui Yang[4] and Xiaojia Wang[1,\*]*

[1]Department of Mechanical Engineering, University of Minnesota, Minneapolis, MN 55455, USA

[2]Department of Electrical and Computer Engineering, University of Minnesota, Minneapolis, MN 55455, USA

[3]Department of Physics and Astronomy, Stony Brook University, Stony Brook, NY 11794-3800, USA

[4]Department of Mechanical Engineering, University of Colorado at Boulder, Boulder, CO 80309-0427, USA

[5]Institute of Engineering Thermophysics, Chinese Academy of Sciences, Beijing 100190, China

[6]Key Laboratory of Thermo-Fluid Science and Engineering of MOE, Xi'an Jiaotong University, Xi'an, Shaanxi 710049, China

[\*]Authors to whom correspondence should be addressed: <u>wang4940@umn.edu</u>

**Black phosphorus (BP) has emerged as a direct-bandgap semiconducting material with great application potentials in electronics, photonics, and energy conversion. Experimental characterization of the anisotropic thermal properties of BP, however, is extremely challenging due to the lack of reliable and accurate measurement techniques to characterize anisotropic samples that are micrometers in size. Here, we report measurement results of the anisotropic thermal conductivity of bulk BP along three primary crystalline orientations, using the novel time-resolved magneto-optical Kerr effect (TR-MOKE) with enhanced measurement sensitivities. Two-dimensional beam-offset TR-MOKE signals from BP flakes yield the thermal conductivity along the zigzag crystalline direction to be 84 ~ 101 W m$^{−1}$ K$^{−1}$, nearly three times as large as that along the armchair direction (26 ~**

**36 W m$^{-1}$ K$^{-1}$). The through-plane thermal conductivity of BP ranges from 4.3 to 5.5 W m$^{-1}$ K$^{-1}$. The first-principles calculation was performed for the first time to predict the phonon transport in BP both along the in-plane zigzag and armchair directions and along the through-plane direction. This work successfully unveiled the fundamental mechanisms of anisotropic thermal transport along the three crystalline directions in bulk BP, as demonstrated by the excellent agreement between our first-principles-based theoretical predictions and experimental characterizations on the anisotropic thermal conductivities of bulk BP.**

Recently, black phosphorus (BP)[1,2] has drawn increased attention as a functional material because of its distinct electronic[2-5], optical[6-8], and thermal properties[9-18]. BP has an orthorhombic structure with puckered layers of sp$^3$-hybridized phosphorus atoms along the armchair direction and van der Waals interactions between layers[2]. The structure of BP makes it the most thermodynamically stable phosphorus allotrope at ambient temperature and pressure. Both monolayer (also called phosphorene) and few-layer BP flakes obtained from mechanical exfoliation[19,20] have proven to be anisotropic two-dimensional (2D) semiconductors with high carrier mobility[5,19-21], possessing a direct band gap tunable from 0.3 to 1.0 eV[22]. This combination of properties makes BP a great candidate for use in the next generation nano-electronic and nanophotonic devices, compared even favorably to other well-studied 2D materials such as graphene, hexagonal boron nitride, and molybdenum disulfide[23-25]. One unique feature of BP is the inverse dependence of its in-plane electronic and thermal transport on the crystalline orientation: a higher electrical conductivity is associated with a lower thermal conductivity in the armchair direction and vice versa for the zigzag direction, which has been predicted in previous studies[4,9].

To date, experimental investigations on the anisotropic thermal properties of BP mainly focused on thin films; thermal conductivity measurements of bulk BP or 2D monolayer BP remain scarce in literature. One difficulty comes from the challenge in synthesizing large-size high-quality BP crystals. The first experimental investigation of the thermal conductivity of bulk BP even failed to capture the anisotropy in the thermal transport of BP[26]. Another impediment stems from the lack of reliable and accurate experimental techniques to measure the anisotropic thermal properties of thin-film samples. The experimental methods for thermal characterization of monolayer or few-layer 2D materials are even more limited with great uncertainty[27]. Thus, despite the fact that theoretical investigations on the thermal transport in BP have been conducted only for monolayer, the recently reported experimental work by three groups were all for BP thin films (while this manuscript was under review and in revision)[14-16]. More importantly, there are large deviations in these experimental studies: the measured thermal conductivities of BP span from 20 to 86 W m$^{-1}$ K$^{-1}$ for the zigzag direction, and from 10 to 36 W m$^{-1}$ K$^{-1}$ for the armchair direction. The measured in-plane thermal conductivities for BP thin films in Refs. [14] and [16] seem to be dependent on sample thickness. There are no theoretical predictions directly on thin film samples to compare with experimental results. Most theoretical works in literature were trying to predict the thermal conductivity of monolayer BP. However, the trustable theoretical prediction for monolayer BP is 2 or 3 times larger than the experimental results on thin films[9]. Such a discrepancy between theoretical predictions and measurement results has been attributed to the enhanced surface scattering, possible surface contamination due to adsorbates and oxidation, or even sample damage during the fabrication process.

On the other hand, the wide range of the theoretically predicted values calls for experimental validation. The predicted in-plane thermal conductivities of monolayer BP in literature range from 110 to 30 W m$^{-1}$ K$^{-1}$ for the armchair direction, and from 36 to 14 W m$^{-1}$ K$^{-1}$ for the zigzag direction, respectively. Such large deviations are presumably due to the different theoretical models employed to calculate the phonon relaxation time and the thermal conductivity, as well as the details of first-principles calculations. Apparently there is a lack of systematic comparison between theoretical predictions and experimental measurements. There are no available measurement data to validate the wide range of theoretically predicted thermal conductivity values of monolayer BP. The high-fidelity theoretical predictions for the recently measured BP thin films can be challenging.

In this study, we focus on the anisotropic thermal transport in high-quality bulk BP along both the basal-plane and through-plane directions. This serves as the best and unique platform to have a high-fidelity integrated theoretical and experimental investigation, which is of critical importance to reveal the fundamental mechanisms of the anisotropic thermal transport in BP. To achieve this objective, we synthesize large-size high-quality BP crystals and employ the technique of time-resolved magneto-optical Kerr effect (TR-MOKE)[28] to study their anisotropic three-dimensional (3D) thermal conductivities. We avoided the surface contamination and degradation issues of BP by in-situ surface cleaning with ion-milling and encapsulation of the sample with a magnetic terbium iron (TbFe) transducer. With the improved measurement sensitivity as compared with the standard time-domain thermoreflectance, the TR-MOKE technique with the TbFe magnetic transducer allows the in-plane thermal conductivities of BP to be accurately determined with a two-dimensional in-plane beam offset approach. First-principles-based calculations have been performed, for the first time, on bulk BP crystals to reveal the characteristics of thermal transport in BP along both the in-plane and through-plane directions. Excellent agreements have been achieved between theoretical and experimental investigations regarding the thermal conductivities of BP along all three crystalline orientations.

**Results**

**BP synthesis and structure characterization**. Single crystalline BP flake samples were synthesized using a low-pressure synthesis technique[29]. Figure 1A depicts the layered structure BP, with three crystalline directions denoted as *a* (zigzag), *b* (interlayer or though-plane), and *c* (armchair). The BP crystals grown by the process are ribbon-like with the long side of the crystal flake along the zigzag direction, as illustrated in Figure 1B, suggesting a well-defined lattice orientation with respect to the crystal morphology.

The lattice constants and crystalline orientation were determined with X-ray diffraction (XRD, Bruker D8 Discover thin film X-ray diffractometer). To enhance the diffraction signal, ten ribbons were placed on top of a microscope glass slide (amorphous with no XRD feature), parallel to each other along the length direction. A $\theta$-$2\theta$ scan clearly shows a strong diffraction peak corresponding to [0m0] (m = 2, 4, 6) planes, indicating that these crystalline ribbons are placed with the phosphorene layers parallel to the substrate (Figure 1C). The lattice constant is determined to be $b = 10.5$ Å along the interlayer (through-plane) direction, consistent with literature values[30,31]. Rotating the ribbons around the *b* axis, a $\phi$-scan of the [061] planes revealed diffraction peaks centered along the width direction ($\phi = 0°$, $180°$), as illustrated in Figure 1D. The small angular spreading in Figure 1D can be attributed to the slight misalignment of BP ribbons. This strongly suggests a definitive correlation between the shape of ribbon-like BP flakes and the lattice orientation: the narrow side of BP flakes is along the "wrinkle" armchair direction (*c* axis), and the length direction of the ribbon-like flakes corresponds to the zigzag direction (*a* axis), as denoted in Figure 1B. Details about the characterization of sample orientation are presented in Section S1.1 of the Supplementary Information (SI). SEM was used to characterize the dimensions of pristine BP flakes (the right inset of Figure 1B). The thicknesses of the flakes are 30~50 $\mu$m, which suggests that the BP flakes are essentially thermally opaque in our thermal conductivity measurements.

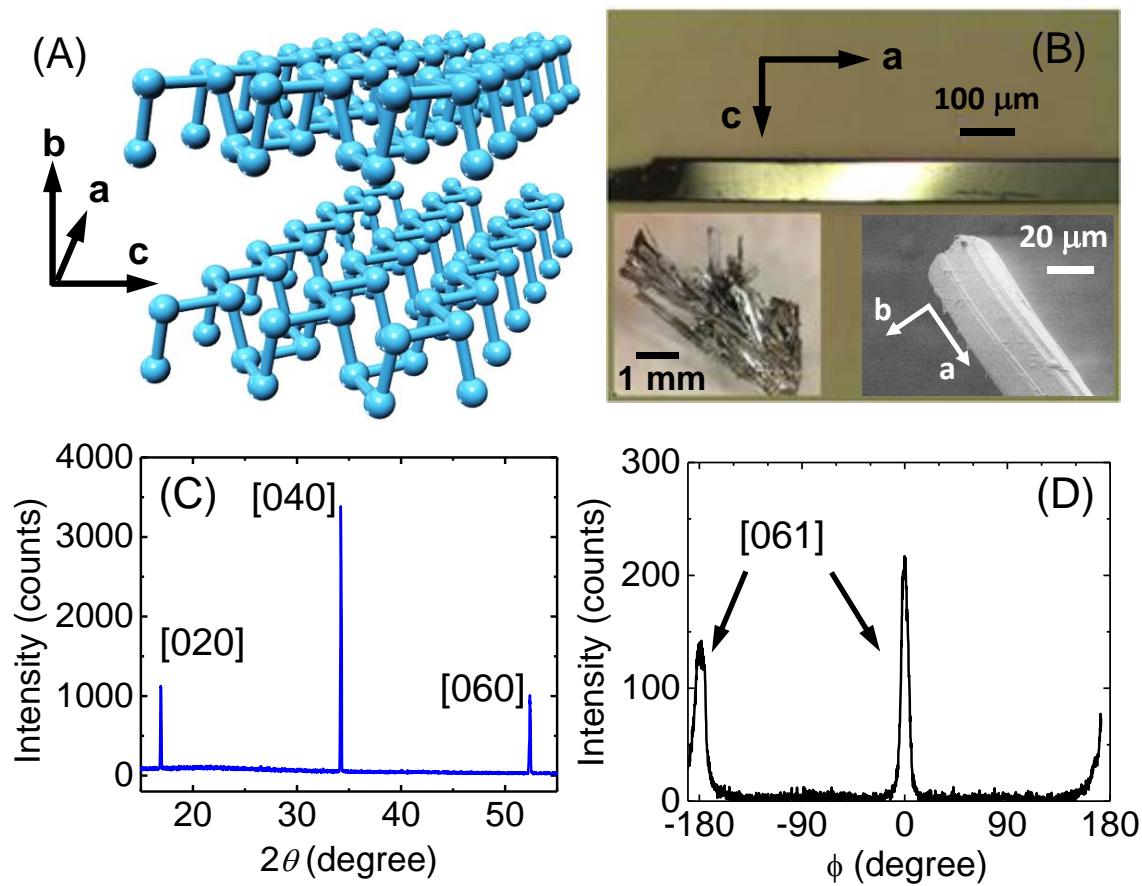

**Figure 1.** (A) Schematic of the lattice structure of BP. The coordinates of *a*, *b*, and *c* correspond to the in-plane zigzag, through-plane, and in-plane armchair directions, respectively; (B) Optical microscopic image of a BP flake that is ribbon-like in shape. Insets: a cluster of BP crystals directly extracted from the growth tube (left) and a SEM image of a BP flake with a thickness ~ 30 $\mu$m (right); (C) XRD $\theta$-$2\theta$ scan showing peaks of [0m0] planes; (D) XRD $\phi$-scan of the [061] planes. Here 0° and 180° are aligned with the width direction of the BP ribbons.

**Through-plane thermal conductivity measurements of BP flakes**. Four different BP flake samples have been measured in this work. BP-1, BP-2 and BP-3 were deposited with TbFe transducers for TR-MOKE thermal-property measurement[28], and BP-4 was coated with an 80-nm Al transducer for through-plane thermal conductivity measurement with time domain thermoreflectance method (TDTR).

In a typical TR-MOKE measurement, in-phase ($V_{in}$) and out-of-phase ($V_{out}$) signals are collected by a balanced photodetector and then an rf lock-in amplifier. Prior to each TR-MOKE measurement, the TbFe transducer is magnetized with an external magnet. The magnetization orientation of TbFe can be flipped by using different poles of the external magnet. Each sample is measured twice with the initial magnetization orientation of TbFe and its reverse, namely, $M+$ and $M-$. Both $V_{in}$ and $V_{out}$ change their signs when the initial magnetization of TbFe is flipped from $M+$ to $M-$. The data used for thermal analysis is the corrected in-phase signal taken as the difference of the two measurements, which is defined as $V_{in} = V_{in}^{M+} - V_{in}^{M-}$. Details about TR-MOKE measurements on both the through-plane and in-plane thermal conductivity of BP samples are provided in Section S2 of the SI.

The TR-MOKE data are fitted with a thermal model to extract the thermal conductivities. In our thermal model, we used the heat capacity ($C_m$), thickness ($h_m$), and thermal conductivity ($\Lambda_{r,m}$ and $\Lambda_{z,m}$) of the metallic transducer (either Al or TbFe), and the through-plane thermal conductivity of BP ($\Lambda_{z,BP}$) as input parameters. Here, the subscript "$z$" denotes the through-plane direction, while "$r$" denotes the in-plane radial direction. $\Lambda_{z,BP}$ also equals to $\Lambda_b$ for BP following the crystalline orientation notation as defined in Figure 1A. To reduce the uncertainties propagated from these input parameters and to validate the experimental approach, we carried out a series of control measurements on the reference samples of $SiO_2$ and obtained the thermal properties of TbFe transducer layers (Section S2.4 of the SI). Figure 2A shows the representative TR-MOKE in-phase signals ($V_{in}$) of sample BP-1 for the through-plane thermal conductivity measurements, taken with $M+$ (red open circles), and $M-$ (black open circles). The data used for thermal analysis is taken as the difference of the two measurements, which effectively exclude any contribution from signals that are not associated with MOKE (Section S2.4 of the SI)[28].

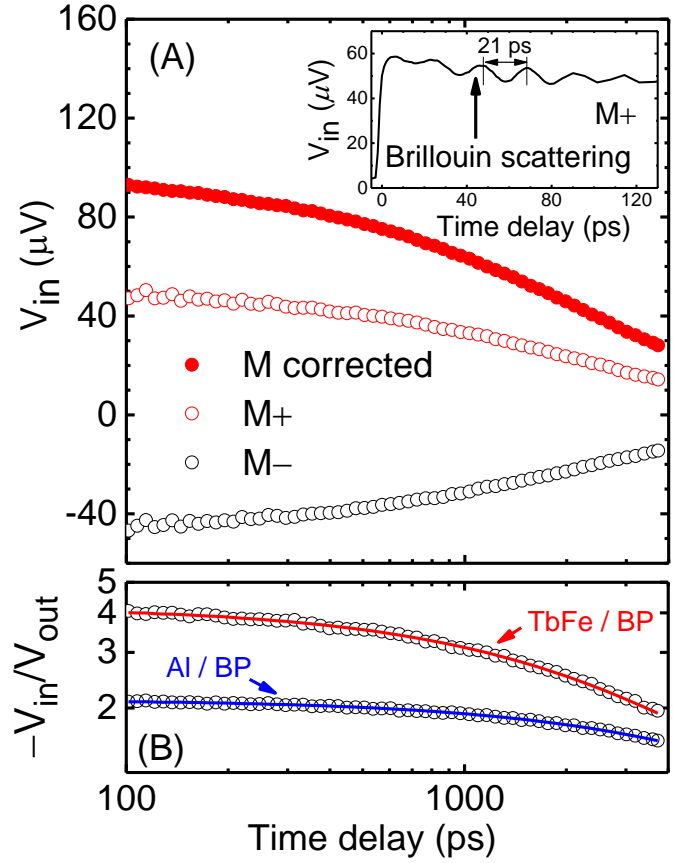

**Figure 2.** Representative TR-MOKE signals from measuring the through-plane thermal conductivity of BP-1 flake. (A) Positive ($M+$, red open circles), negative ($M-$, black open circles) and corrected (red filled circles) $V_{in}$ signals of the through-plane thermal conductivity measurement. The inset plot shows the Brillouin scattering oscillations with a period of ~21 ps occurring at the first several hundreds of ps. These oscillations can be cancelled out in the corrected $V_{in}$ signals ($M$ corrected); (B) The ratio signals of $-V_{in}/V_{out}$ (open circles) for BP flakes of both TDTR measurement with an 81-nm Al transducer (blue) and TR-MOKE measurement with a 26.9-nm TbFe transducer (red), and their thermal model fitting (solid lines).

Figure 2 depicts a representative TR-MOKE data from sample BP-1, measured at 9 MHz with $w_0 = 12$ $\mu$m. Fitting the ratio signals ($-V_{in}/V_{out}$) of the through-plane TR-MOKE measurements, the through-plane thermal conductivity of BP-1 ($\Lambda_{z,BP}$ or $\Lambda_b$) and the interfacial thermal conductance between TbFe and BP-1 are obtained as $\Lambda_b = 4.5 \pm 0.4$ W m$^{-1}$ K$^{-1}$ and $G = 39 \pm 3$ MW m$^{-2}$ K$^{-1}$, respectively. Using the same method on BP-2, we obtained $\Lambda_b = 5 \pm 0.5$ W m$^{-1}$ K$^{-1}$ and $G = 40 \pm 3$ MW m$^{-2}$ K$^{-1}$. Standard TDTR measurement was also performed on both reference (control) samples and BP-4, to compare with the TR-MOKE measurement results. TDTR measurements of BP-4 flake give a result of $\Lambda_b = 5.5 \pm 0.5$ W m$^{-1}$ K$^{-1}$. We attribute this discrepancy in $\Lambda_b$ of BP to the sample variation and/or location dependence. All the four BP flakes gives a thermal conductivity $\Lambda_b$ ranging from 4.3 to 5.5 W m$^{-1}$ K$^{-1}$ with a typical relative uncertainty from 8 to 10%. This lower through-plane thermal conductivity of BP as compared with its in-plane thermal conductivities (30 W m$^{-1}$ K$^{-1}$ < $\Lambda_a$ or $\Lambda_c$ < 100 W m$^{-1}$ K$^{-1}$) is expected given the much weaker van der Waals interlayer interaction along the through-plane direction.

The inset of Figure 2A shows observable oscillations in both $M+$ and $M-$ signals, which we attributed to Brillouin scattering. The Brillouin scattering oscillations have a period of $\tau = 21$ ps, and last for 500 ps of time delay. Using the literature value for the speed of sound along the though-plane direction (or $b$ as the interlayer direction for BP) of $v_b \approx 5100$ m s$^{-1}$ [32], the refractive index $n_b$ of BP along through-plane direction can be calculated as $n_b = \lambda/(2\tau v_b)$ [33], where $\lambda = 783$ nm is the wavelength of the probe beam. The calculation gives $n_b \approx 3.7$ as the real component of the refractive index of BP at 783 nm. We have not found any literature reporting the refractive index of BP at this wavelength; however, we note that this value is quite close to that of silicon at the same wavelength.

The Brillouin scattering oscillations only appear in TR-MOKE with thinner magnetic transducers, which are optically semitransparent. This allows the laser beam to directly excite the BP sample beneath the transducer. In this situation, the signal collected by the balanced photodetector also contains portions of the thermoreflectance and/or Brillouin scattering signals originated from the interference of the probe beam reflected by the moving grating of acoustic wave travelling inside BP with the rest of the interfaces within the sample stack[33]. This acoustic wave is launched by sudden heating of the thin metal film and propagates into the BP. Since subtracting measurement signals of $M+$ and $M-$ removes the contribution from signals that do not generate the Kerr rotation within the magnetic layer, these oscillations are cancelled out by the corrective subtraction. The appearance of this phenomenon also verifies the assumption of TR-MOKE that the part of signals which are not from Kerr effect can be removed by subtracting the difference in the signals for positive and negative magnetization. The Brillouin scattering oscillations are missing in TDTR with the optically thick Al transducer (≈80 nm).

**In-plane measurements of BP by two-dimensional (2D) beam offset**. To map out the anisotropy in the in-plane thermal conductivity of BP, we conducted 2D TR-MOKE by in-plane scanning using the beam-offset approach[34,35]. Results from this 2D-scanning method generate a contour plot of in-plane thermal conductivity of BP, which allows the direct extraction of the in-plane thermal conductivity along a primary crystalline direction accurately. This method does not require a precise sample alignment with a specific crystalline orientation during the loading process, and therefore greatly reduces the measurement difficulties in thermal characterization of anisotropic samples that are a few microns in size.

The contour plots of out-of-phase signals ($V_{out}$) measured by 2D TR-MOKE are illustrated in Figures 3A and 3B, for BP-1 loaded at 0° and 90° orientation angles, respectively. The $X$ axis of the beam offset direction is along the zigzag orientation of the BP flake when the sample is loaded at 0°, but along the armchair orientation if loaded at 90°. Corrections on $V_{out}$ mapping signals are made (Section S2.7 of the SI), which are essential for processing the measurement data to obtain accurate thermal properties of the sample.

Figures 3C and 3D are the beam-offset signals along $X$ and $Y$ directions (white solid lines) from the 2D contour plots in Figures 3A and 3B. The values of the full width at half maximum (FWHM) of measured $V_{out}$ are compared with those predicted from the thermal model to obtain the in-plane thermal conductivity of BP along either $X$ or $Y$ directions. The in-plane thermal conductivity of the TbFe transducer (17.4 W m$^{-1}$ K$^{-1}$) and the through-plane thermal conductivity of the BP flake (4.5 W m$^{-1}$ K$^{-1}$) are used for FWHM fitting. These values are obtained from the in-plane measurement of the reference SiO$_2$ sample and the through-plane measurement of the BP-1 sample with the TbFe transducer. The in-plane thermal conductivities of BP-1 loaded at 0° are found to be 91 ± 10 W m$^{-1}$ K$^{-1}$ along the $X$ direction and 26 ± 3 W m$^{-1}$ K$^{-1}$ along the $Y$ direction, corresponding to the zigzag and armchair crystalline directions, respectively (Figure 3C). When the BP flake is loaded at 90°, the model fitting gives 31 ± 3 W m$^{-1}$ K$^{-1}$ along $X$ (armchair) and 84 ± 9 W m$^{-1}$ K$^{-1}$ along $Y$ (zigzag), as seen in Figure 3D. The averaged in-plane thermal conductivities over the two measurements on both orientations of BP-1 are 87.5 ± 10 W m$^{-1}$ K$^{-1}$ (zigzag) and 28.5 ± 4 W m$^{-1}$ K$^{-1}$ (armchair), respectively. The anisotropic factor of in-plane thermal conductivity $\eta = \Lambda_a/\Lambda_c = 3.1$ is in excellent agreement with our first-principles-based theoretical prediction of bulk BP as detailed later in the discussion section.

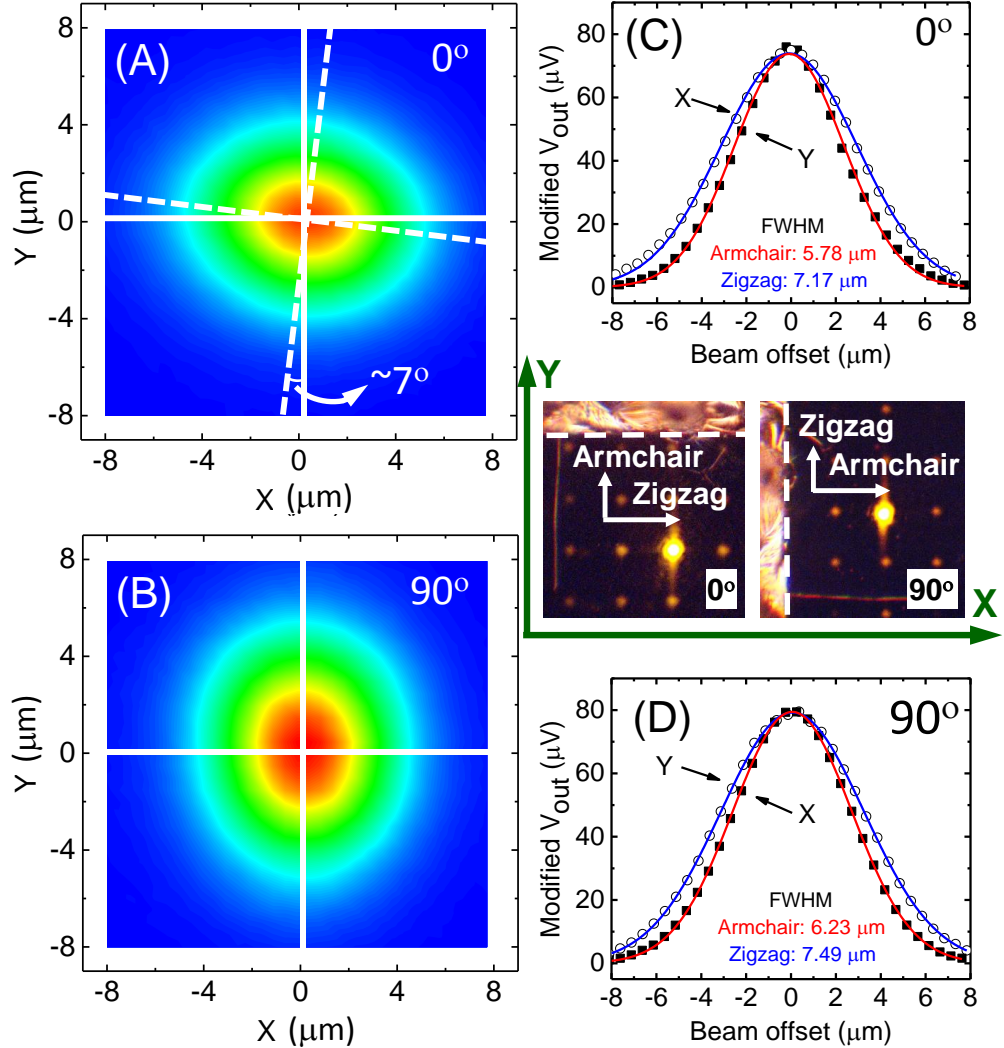

**Figure 3.** In-plane beam offset contour measured on BP-1 at (A) 0° of sample-loading orientation and (B) 90° of sample-loading orientation. For both orientations, measurements were conducted with the 20× objective lens and a 1.6 MHz laser modulation frequency to facilitate the in-plane thermal transport. The two beam spot sizes measured at 20 ps time delay are 2.9 μm and 3.1 μm in (A) and (B). The beam offset signals in (C) and (D) are extracted from (A) and (B), respectively, along the *X* and *Y* directions. The FWHM of $V_{out}$ signals in zigzag (open circles) and armchair

(filled squares) directions are fitted with the heat transfer model to determine the in-plane thermal conductivity of BP.

In Figure 3A, it is clearly observed that there exists a small offset angle between the primary axes of the elliptical contour and the mapping scanning directions of $X$ and $Y$, indicating that the sample loading orientation does not perfectly match the crystalline orientations of BP; this could potentially introduce deviations to measurement results. To examine how large these deviations could be, we pick the actual primary axes of the contour and get a pair of new FWHM of $V_{out}$ as 7.23 $\mu$m and 5.76 $\mu$m. The angle between the actual primary axes of the contour and the axes of $X$ and $Y$ is calculated to be ~7°. Compared with the values taken along $X$ and $Y$ (7.17 $\mu$m and 5.78 $\mu$m), this deviation generates < 1% uncertainty in the FWHM of $V_{out}$, which suggests that a small imperfection of sample loading orientation (< 7°) does not significantly affect the in-plane thermal conductivity measurement results.

We then check the dependence of thermal conductivity on sample variation by making beam-offset measurements on BP-2 with both 0° and 90° sample loading orientations. The in-plane thermal conductivities of BP-2 are 101 ± 10 W m$^{-1}$ K$^{-1}$ and 36 ± 4 W m$^{-1}$ K$^{-1}$, respectively, for the zigzag and armchair directions measured at 0°, while 96 ± 10 W m$^{-1}$ K$^{-1}$ and 33 ± 4 W m$^{-1}$ K$^{-1}$ at 90°. The averaged in-plane thermal conductivities over the two measurements on both orientations of BP-2 are 98.5 ± 10 W m$^{-1}$ K$^{-1}$ (zigzag) and 34.5 ± 4 W m$^{-1}$ K$^{-1}$ (armchair). The anisotropic factor of in-plane thermal conductivity $\eta$ = 2.9. Due to the short depth of focus of the high-power objective lens, the laser beam spot used in the BP in-plane measurements varies from set to set in the range of 2.9 to 3.2 $\mu$m for the 20× objective lens. However, with careful beam spot shaping, the ellipticity of the laser beam spots are nearly 1 ± 0.01, suggesting that the measured anisotropy of BP in-plane thermal conductivity contains negligible contribution from the shape of the laser spot (Section S2.2 of the SI).

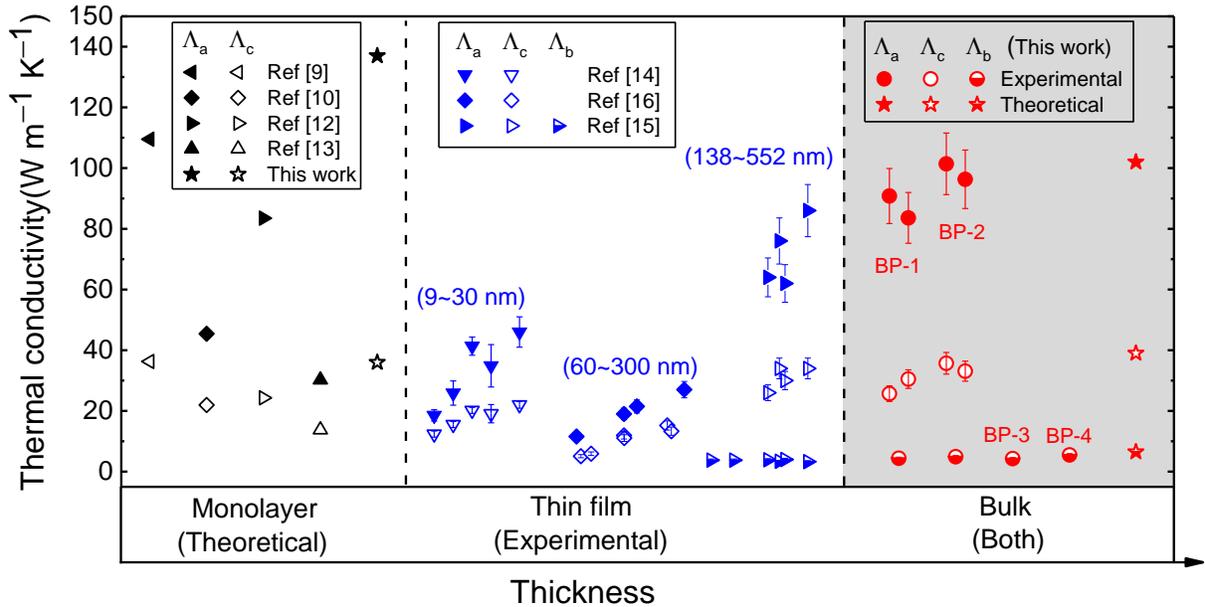

**Figure 4.** Summary of the anisotropic thermal conductivities: highlighted in gray are for bulk and BP from both our measurements and simulations at 300 K. The available literature values (theoretical for monolayer BP[9,10,12,13] and experimental for BP thin films[14-16]) are also plotted for comparison.

The thermal conductivities of all the four BP flake samples measured at room temperature and predicted with first-principles calculation are summarized in Figure 4. For comparison, literature values of the thermal conductivities of BP obtained from both theoretical prediction and measurements are also presented. Clearly seen in this figure, most of the theoretical works were on monolayer BP while the experiments have been confined to ad-hoc thin films with a submicron thickness depending on the sample availability. We note here that we did not measure the in-plane thermal conductivity for BP-3 due to the lack of sufficiently large and clean areas for TR-MOKE measurements. Looking into the averaged values obtained on BP-1 and BP-2, the differences of measured in-plane thermal conductivity between different samples are within the range of 12% along both zigzag and armchair directions.

**Discussion**

As demonstrated by the excellent agreement on the thermal conductivity between the first-principles calculation and TR-MOKE measurements, the high-quality bulk BP serves as the best and unique platform to investigate the fundamental mechanisms of the anisotropic thermal transport in this class of layered 2D materials. Our observations clearly indicate that the thermal conductivity of single crystalline BP is strongly anisotropic not only between in-plane (covalent bonding) and through-plane (van der Waals interaction) directions, but also within the in-plane directions (zigzag vs. armchair).

To understand the origins of the anisotropic thermal conductivity in single crystalline bulk BP, we performed the first-principles-based Peierls-Boltzmann transport equation (PBTE) calculations to study phonon transport along different crystalline directions (See calculation details in Section S3 of SI). By solving the PBTE through the iterative approach, the predicted thermal conductivities of bulk BP at room temperature are 102 (zigzag), 6.5 (through-plane), and 39 (armchair) W m$^{-1}$ K$^{-1}$, respectively. As shown in Figure 4, our theoretical calculation is in great agreement with the measured thermal conductivities, which indicates that our theoretical calculations successfully capture the fundamentals of anisotropic phonon transport in BP. We identify that the strong anisotropy of thermal transport in BP comes from two aspects: one is the structural-asymmetry-induced group velocity variations along different crystalline orientations, which can be readily calculated from the phonon dispersion of BP; and the other is the relaxation time variation induced by the direction of the applied temperature gradient.

Previous theoretical works usually rely on the solution of PBTE using the single-mode relaxation time approximation (SMRTA)[11,12,14]. Under the SMRTA, the relaxation process of a specific phonon mode is assumed to be independent to the states of other phonons, which are regarded as staying in their own equilibrium states. The relaxation time of a phonon mode is thus considered to be irrelevant to the direction of the imposed temperature gradient, but an intrinsic property of the phonon mode, which was calculated using the third-order anharmonic force constants from the first-principles calculation. Therefore, the anisotropic group velocity is considered as the only origin for the anisotropic thermal conductivity in these literature works. We calculated the thermal conductivity using PBTE solution with SMRTA and obtained the thermal conductivity of 77, 33, and 5.8 W m$^{-1}$ K$^{-1}$ for the zigzag, armchair and through-plane directions, respectively. Although these thermal conductivity values are apparently anisotropic using SMRTA assumption, these values are quite different from experimental characterization, which are 101, 5, and 36 W m$^{-1}$ K$^{-1}$ for the zigzag, armchair and through-plane directions, respectively. Instead, we performed the exact solution of the PBTE using the iterative approach and obtained the thermal conductivity values of 102, 6.5, and 39 W m$^{-1}$ K$^{-1}$ for the zigzag, armchair and through-plane directions, respectively. While the solution of PBTE with SMRTA gives lower values of thermal conductivity, the exact solution of PBTE using the iterative approach obtained the values agreeing well with the experimental results. This means that the group velocity alone is insufficient to explain the large degree of anisotropy on thermal conductivity, and we have to consider the collective scattering behavior of all the phonon modes beyond the SMRTA. This is because under a temperature gradient all phonon modes are driven to deviate from their equilibrium phonon population to a non-equilibrium one. The scattering of a specific phonon mode is thus correlated to the non-equilibrium population of other phonons. When the direction of temperature gradient is altered, the non-equilibrium phonon population distribution changes correspondingly, leading to different relaxation times, which again depend on the non-equilibrium population of other phonons.

Figure 5A shows a quantitative example on the difference between the phonon relaxation time calculated from the iterative approach with respect to the direction of temperature gradient imposed along the armchair direction and that from the SMRTA. To guide the eyes visually, the black dash curve indicating that the relaxation times from each method equal with each other is plotted in Figure 5A. The relaxation time from the iterative approach can be either larger or smaller than that from the SMRTA, and the ratio between them can be up to hundreds for many phonon modes. It is even counterintuitive to find that for a large portion of phonon modes the relaxation time calculated from the iterative approach can even become negative. Under SMRTA, the relaxation times should be positive. When the iterative approach was used, the negative relaxation time should be understood as that the non-equilibrium populations of those modes are decreased (increased) compared with their equilibrium population, if the phonon mode has a positive (negative) group velocity component along the direction of temperature drop. To observe how the relaxation time changes with the direction of temperature gradient imposed, we show that the relaxation time as a function of phonon frequency in Figure 5B. It is evident that the relaxation time becomes quite different when the direction of temperature gradient is changed.

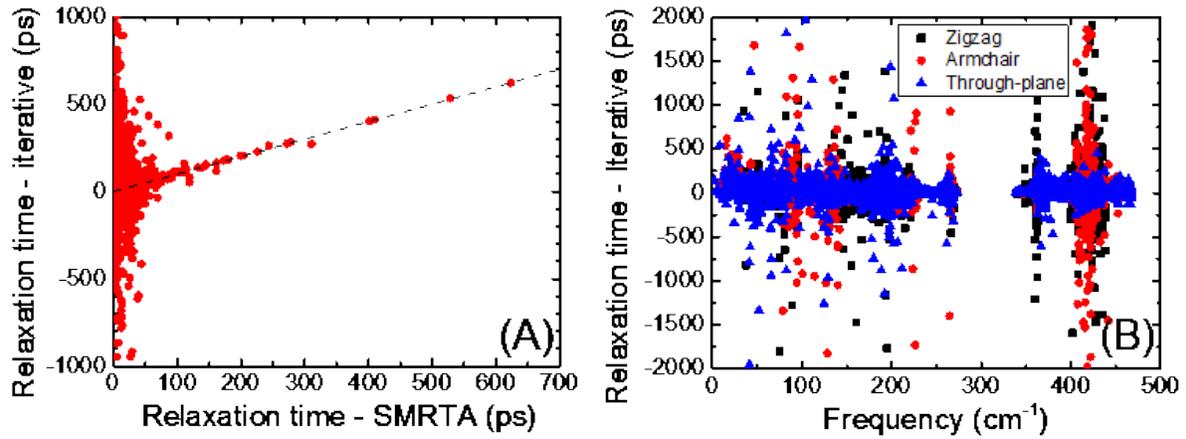

**Figure 5**. (A) The relation between phonon relaxation time from the SMRTA and that from the iterative approach when the temperature gradient is applied along the armchair (*c*) direction. (B) The relaxation time from the iterative approach as a function of phonon frequency when the temperature gradient is applied along different directions.

Figure 6 shows the temperature-dependent thermal conductivity of single crystalline bulk BP. The van der Waals solids have a much weaker phonon-phonon scattering than the conventional 3D solids. As a results, the thermal conductivity in all directions decreases gradually with the temperature.

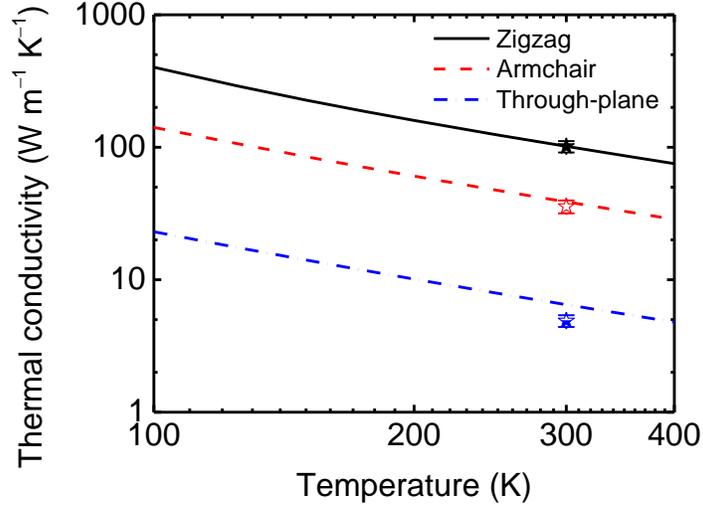

**Figure 6**. Temperature-dependent thermal conductivity of single crystalline bulk BP calculated from first-principles-based PBTE method. Symbols are the experimental thermal conductivity values obtained for BP-2 at room temperature in this work.

In summary, we studied the anisotropic thermal conductivities of single crystalline BP flakes along the three primary crystalline orientations using ultrafast laser-based TR-MOKE beam offset techniques. We successfully validated the measurement results by comparing them with first-principles calculation on bulk BP, as demonstrated by the excellent agreements on the thermal conductivities of BP along both the in-plane and through-plane directions. The through-plane thermal conductivity of BP was measured by TR-MOKE with 0° orientation and validated with the conventional TDTR method. A strong Brillouin scattering feature has been found in the signals with a period around 21 ps during 0 to 500 ps time delay. Two-dimensional contours of $V_{out}$ signals were obtained in the beam-offset measurement and the FWHM of $V_{out}$ were compared against a heat transfer model to fit for the in-plane zigzag and armchair thermal conductivity values. The 2D-scanning method paves a new way of determining the crystalline orientations of BP flakes not only more accurately, but more easily as well. This is usually one of the challenges in thermal characterization of anisotropic materials at the microscale. The measured through-plane thermal conductivities of the BP samples range from 4.3 to 5.5 W m$^{-1}$ K$^{-1}$ at room temperature. The in-plane thermal conductivities of the BP samples range from 84 to 101 W m$^{-1}$ K$^{-1}$ and 26 to 36 W m$^{-1}$ K$^{-1}$ along the zigzag and armchair orientations, respectively, leading to an anisotropic factor of $\eta \approx 3$ for in-plane thermal conductivity. The work not only provides a better understanding of the fundamental mechanisms of anisotropic thermal transport in BP, but also enables potential applications of BP in nanoelectronics, nanophotonics, or other energy related devices where thermal management is crucial.

**Methods**

**Sample preparation.** For samples of BP-1, BP-2, and BP-4 (from the Stony Brook University), a mixture of 4.7 mg of Sn, 2.4 mg of SnI4, and 117.6 mg of red phosphorus are vacuum sealed into a quartz ampoule with an inner diameter of 4.75 mm and length of 10 cm. The sealed quartz ampoule is placed inside a tube furnace and heated to 650°C. After maintaining the temperature at 650°C for one hour, the furnace is cooled at a constant rate of 0.22°C per minute to 300°C over a long duration of 26.2 hours, during which the BP crystals are formed. Finally the quartz ampoule is cooled down to room temperature and the BP crystals are extracted by mechanically breaking the ampoule. For BP-3 (from the University of Minnesota), a similar process by loading a grain powder consisting of 12 mg of SnI4, 22 mg of Sn, and 501 mg of amorphous red phosphorus into one end of a quartz ampoule. The end of the ampoule containing the powder is heated to 650°C, while the sealed end reaches 600°C. The ampoule is maintained at these conditions for one hour and then cooled to room temperature over a period of six hours. The process produces a coating of red phosphorus on the wall of the ampoule near the cold end, followed by a thin layer of amorphous BP, from which the crystalline BP grows based on the growth mechanism of mineralization as proposed by Nilges[30].

BP is known to experience irreversible degradation when exposed to ambient conditions, due to the chemical reaction with ambient oxygen[36]. To prevent sample degradation, the quartz ampoule is placed in a low-oxygen glove box when the BP crystals are extracted from the ampoule. Right after the mechanical breaking of the ampoule, the BP crystals are picked up and encapsulated in a vacuum vial prior to mechanical exfoliation and the thin-film deposition of metallic transducers.

**Deposition and characterization of metal transducer.** The TbFe magnetic transducer layer with a thickness of ~26.9 nm for TR-MOKE was prepared by co-sputtering of Tb and Fe targets whose composition was controlled by varying the dc power applied to each target. The thickness of the TbFe layer was measured using X-ray reflectivity. The TbFe magnetic properties were characterized by using a vibrating sample magnetometer, and an excellent perpendicular magnetic anisotropy with almost 100% remanence was observed. The saturated magnetization (Ms) is ≈420 emu cm$^{-3}$ and the coercivity (Hc) is ≈1.97 kOe (Figure S3A in the SI). The composition of TbFe was determined by Rutherford backscattering spectroscopy (Figure S3B in the SI). For TDTR measurements, a thin film of Al (≈80 nm) was sputtered onto the BP sample along with the reference samples of SiO$_2$ and Si. The thermal conductivity of Al was derived based on its electrical conductivity measured with a standard four-point probe station and the Wiedemann-Franz law. The thickness of the Al transducer was measured from picosecond acoustics[37].

**Thermal conductivity measurement**. Both TDTR and TR-MOKE are noncontact methods based on a pump-probe configuration, in which a mode-locked Ti:sapphire laser produces a train of optical pulses (~100 fs in duration) at a repetition rate of 80 MHz. A polarizing beamsplitter divides the laser into a pump beam and a probe beam with two orthogonal polarizations. A mechanical delay stage is used to vary the relative optical path length between the pump and probe before they are focused on the sample surface through a single objective lens. The pump beam, modulated by an electro-optical modulator, heats the sample. The probe signals are measured by an rf lock-in amplifier in both TDTR and TR-MOKE measurements. All the thermal conductivity measurements in this work were taken at room temperature.

For the through-plane thermal conductivity measurements of the reference SiO$_2$ sample and BP flakes, a 5× objective lens was used to produce a $1/e^2$ spot radius of $w_0 = 12$ $\mu$m for both pump and probe beams. The pump light was modulated at frequency of 9 MHz. The total laser power was chosen to be 18 mW with a power ratio of 2:1 for pump and probe beams, which provides an excellent balance between a good signal-to-noise ratio and a moderate steady-state heating at room temperature (Section S2.4 of the SI) for all samples studied in this work. For in-plane thermal measurements of the reference samples and BP flakes, a 20× objective lens with a $1/e^2$ spot radius of $w_0 = 3$ $\mu$m and a modulation frequency of 1.6 MHz were used. All in-plane thermal conductivity measurements were taken at negative time delay of −50 ps. Details about TDTR and TR-MOKE measurements are provided in Section S2 of the SI.

**Data analysis**. Procedures for modeling of TDTR and TR-MOKE measurements have been discussed in detail previously[38-40]. The ratio of the in-phase to out-of-phase signal ($-V_{in}/V_{out}$) is used to analyze the response which improves the signal-to-noise ratio, increases the sensitivity, and minimizes artifacts created by variations of laser spot size and beam overlap as a function of delay time[38]. The thicknesses of pristine BP flakes are on the order of tens of microns as characterized by scanning electron microscopy. There were no observable acoustic echoes from the buried BP samples/tape interface which suggests that the BP flakes can be treated as thermally opaque, which qualifies them to be treated as bulk samples. For accurate thermal modeling of the experimental data, we analyzed the measurement sensitivity to different parameters (Section S2.3 of the SI).

**First-Principles Calculation.** When temperature gradient is imposed on a material, the phonon population of each phonon mode changes from the equilibrium Bose-Einstein distribution to an non-equilibrium one. We solved the Peierls-Boltzmann transport equation (PBTE), which describes the balance between the phonon diffusion driven by the temperature gradient and the phonon scattering, to obtain the non-equilibrium phonon population. Details of the PBTE, including its formalism, the iterative solution of PBTE and the single-mode relaxation time approximation (SMRTA) of PBTE, are presented in Section S3.1 of the SI. In the PBTE, phonon dynamics information, such as phonon dispersion and phonon scattering rate, is needed as inputs. We employed the density-function theory (DFT) calculations to extract the second-order harmonic and third-order anharmonic force constants of black phosphorus to calculate the phonon dispersion and phonon scattering rates. The procedures of DFT-based first-principles calculations and extracting the force constants are summarized in Section S3.2 of the SI.

# Supplementary information

## Table of contents

# S1. Sample characterizations

**S1.1 XRD characterization to correlate the sample directions and crystalline orientations**

To correlate the ribbon-like sample shape and the crystalline orientations using X-ray diffraction (XRD), the [0m0] planes and the direction of the $b$ axis were first determined by aligning the crystal in the X-ray diffractometer in order to optimize the intensity of the [0m0] diffraction peaks (Figure 1C). To indentify the in-plane lattice orientation, we chose to measure the [061] planes which have relatively strong diffraction peaks (as depicted in Figure 1D) and an in-plane component parallel to the $c$ axis (armchair direction). The direction of the [061] planes was described by the vector perpendicular to the planes, as indicated by the solid red line in Figure S6. To indentify the [061] peaks, both the direction ($\alpha$, in relative to the $b$ axis) and the spacing ($d$) of the [061] planes were calculated using the lattice constants of BP. In XRD measurements, both the X-ray beam and detector were tilted so that 1) the diffraction angle $\theta$ satisfies the Bragg relation ( $2d\sin\theta = n\lambda$ ); 2) the angle bisector of the incident beam and the detector is pointing to the [061] direction. With this approach, the diffractometer was "locked" to the [061] planes. By rotating the sample along the *b* axis, a strong diffraction peak appeared when the length of the BP ribbons was perpendicular to the plane of the incidence for the X-ray beam. This means that the narrow side of the ribbon-like BP flakes is along the *c* axis, hence along the armchair direction. And the length direction of the ribbon-like BP flakes corresponds to the zigzag direction (the *a* axis).

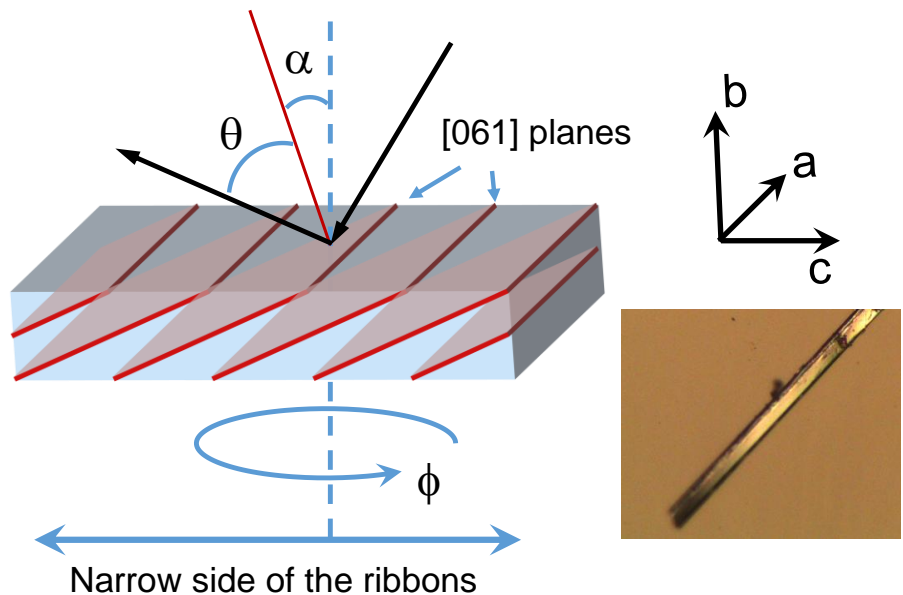

**Figure S1.** X-ray diffraction measurements for corresponding between sample directions and crystalline orientations.

**S1.2 Sample characterization for quality check**

Defect characterization is not a simple matter for BP flakes. The most sensitive techniques are electrical characterization, including deep level transient spectroscopy (DLTS), photocapacitance, and impedance spectroscopy. However, all these three electrical measurements require at a minimum the formation of lateral p-n junctions. This is extremely difficult for BP because of our limited understanding of its doping. We have used Raman spectroscopy and high-resolution transmission electron microscope (HR-TEM) to inspect BP samples as shown in Figure S2, in addition to the XRD characterization presented in Figures 1C and 1D. The peaks in both Raman and XRD are sharp and all of the peaks can be indexed to BP. The HR-TEM results clearly show the expected BP structure. Although these characterizations cannot completely rule out the presence of defects, they indicate that the BP samples are highly crystalline and in good quality.

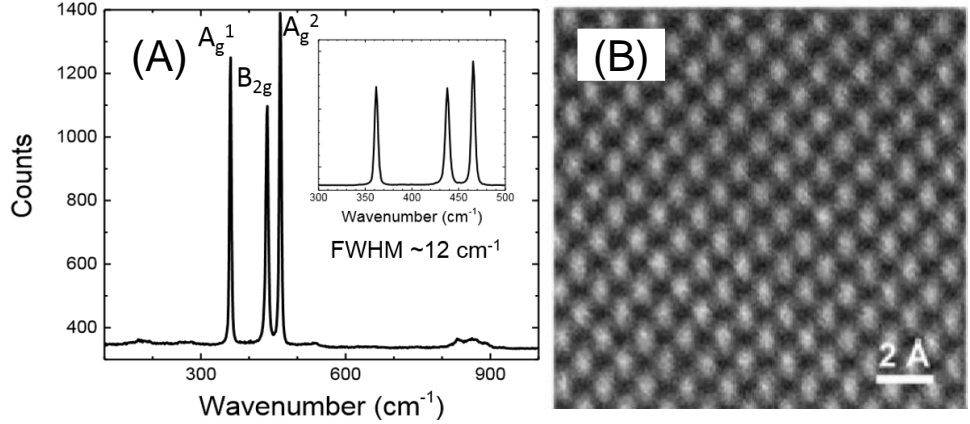

**Figure S2.** (A) Raman and (B) high-resolution transmission electron microscopy characterization of the crystalline BP-3 flake.

## S2. Thermal conductivity characterization: TDTR and TR-MOKE

The anisotropic thermal conductivities of the BP samples are characterized by the ultrafast laser-based TR-MOKE[1]. In TR-MOKE, a balanced photodetector senses the polarization variation of the probe beam reflected from the magnetic transducer. This polarization variation is linearly correlated to the thermal demagnetization-induced Kerr rotation angle of the magnetic transducer ($\theta_k$), when it is subjected to a small temperature rise upon pump heating and a subsequent cooling[1]. Comparing to the standard TDTR which monitors the temperature-induced reflectivity change in the probe beam reflected from an optically thick (~70-100 nm) metallic film[2]. TR-MOKE allows for the use of optically thin magnetic transducers (~20-30 nm) with relatively low thermal conductivities and enhances the measurement sensitivity, especially for the in-plane thermal conductivity characterization of the underlying material[1-3]. The sensitivity analysis also confirms that TR-MOKE serves as a better choice for the in-plane thermal conductivity characterization of BP.

For the through-plane thermal conductivity measurements of the reference $SiO_2$ sample and BP flakes, a 5× objective lens was used to produce a $1/e^2$ spot radius of $w_0 = 12\ \mu m$ for both pump and probe beams. A modulation frequency of 9 MHz was used for the through-plane thermal conductivity measurements. For through-plane measurements of anisotropic BP flakes, the combination of large beam spot size and high modulation frequencies facilitate one-dimensional heat flow approximation; therefore, the anisotropy has negligible effects on extracting the through-plane thermal conductivity. The total laser power was chosen to be 18 mW with a power ratio of 2:1 for pump and probe beams, which provides an excellent balance between a good signal-to-noise ratio and a moderate steady-state heating at room temperature for all samples studied in this work ($\Delta T_{ss} \approx 10$ K as detailed in Section S2.5).

For the in-plane thermal conductivity measurements of the reference samples and BP flakes, a 20× objective lens with a $1/e^2$ spot radius of $w_0 = 3$ $\mu$m and a laser modulation frequency of 1.6 MHz were used. This combination of a smaller beam spot size and a low modulation frequency facilitates the in-plane heat spreading, which improves the measurement sensitivity to the in-plane thermal transport in BP flakes. All in-plane thermal conductivity measurements were taken at negative time delay of −50 ps to minimize the measurement uncertainty due to errors in setting the reference phase[1,9]. The TR-MOKE measurements were conducted for an area of 16 $\mu$m × 16 $\mu$m to extract in-plane thermal conductivity. Some artifacts in the TR-MOKE signal are expected due to the drift of laser power and wavelength over a long duration of scanning (1 ~ 2 hours). We minimize those artifacts by correcting the TR-MOKE signal with the power signals of both the pump and probe beams monitored with a fast-response power meter and a photodiode detector, throughout the entire measurement duration. This signal correction approach adequately suppresses those artifacts.

**S2.1 Deposition and characterization of metal transducers**

Amorphous rare-earth transition metal (RE-TM) alloys, such as GdFeCo, TbFeCo, and TbFe, have been investigated for magneto-optical recording applications in the past decades[4,5]. This family of materials has a large perpendicular magnetic anisotropy and a strong magneto-optical effect, which generates large Kerr rotation. In particular, the amorphous nature of RE-TM alloys gives relatively low magneto-optical media noise due to optical inhomogeneity, which is an advantage over polycrystalline magnetic films[5]. In our work, a layer of amorphous RE-TM alloy, TbFe, has been used as magnetic transducer layer for thermal conductivity measurement via the time-resolved magneto-optical Kerr effect (TR-MOKE). TbFe is a ferrimagnet and the spins of its Tb and Fe sublattices are antiferromagnetically coupled with each other. The saturated magnetization ($M_s$), coercivity ($H_c$), and the perpendicular magnetic anisotropy can be easily controlled by varying the composition of Tb. The magneto-optical effect of TbFe under 780-nm laser illumination mainly originates from Fe atoms. Its Kerr rotation angle is comparable to that of other magneto-optical materials, such as Co/Pt and Co/Pd multilayers[6]. Amorphous TbFe alloys have low Curie temperatures and poor thermal stability. This is detrimental for recording device applications; however, these same properties result in a large thermal-magnetic variation, which is desirable for enhancing signal-to-noise ratio in our thermal conductivity measurement.

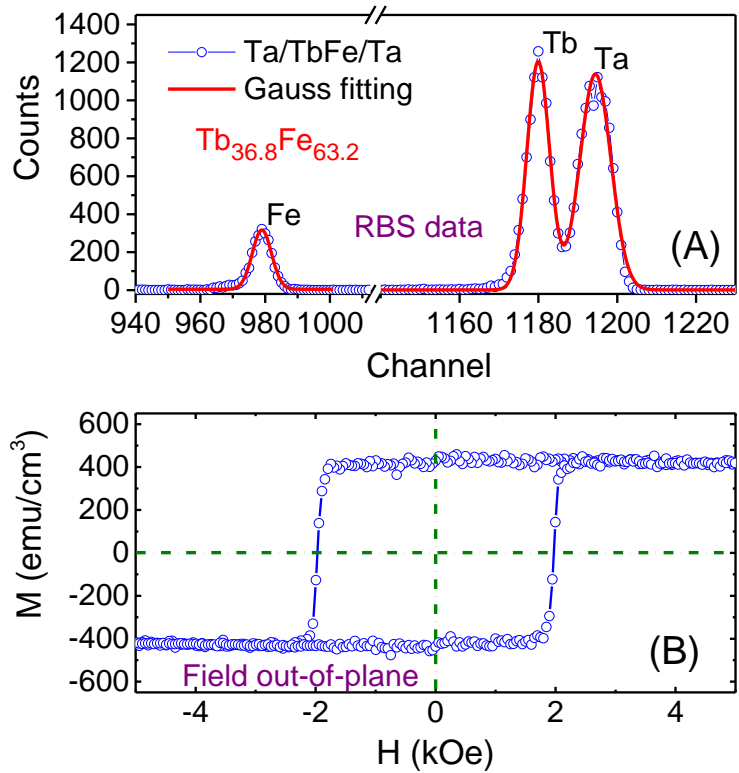

**Figure S3**. Characterization of magnetic TbFe transducer layer. (A) Rutherford backscattering spectroscopy of TbFe layer to determine its composition; (B) Hysteresis loop of TbFe with magnetic field applied along the out-of-plane direction. It shows almost 100% remanance and a coercivity of ~1.97 kOe (1 kOe = 79.6 A m$^{-1}$), which indicates a good perpendicular magnetic anisotropy of the film.

The BP flakes were mechanically exfoliated and loaded onto a conductive vacuum-compatible tape (Thermattach T410) supported by a silicon substrate. The magnetic transducer was sputtered onto two reference samples of (a) 300-nm SiO$_2$ on silicon and (b) bare silicon, together with mechanically exfoliated BP flakes.

The composition, thickness, and magnetic properties of the TbFe transducer layer have been calibrated carefully. The films consist of TbFe (19.7 nm) with Ta (3.6 nm) layers as seed and protection layers. The composition of TbFe film has been checked by fitting the peaks of Tb and Fe in Rutherford Backscattering spectroscopy, which gives the composition of Tb ($x_{Tb}$) ~36.8% as shown in Figure S3A. Figure S3B shows the hysteresis loop of TbFe with the magnetic field applied out-of-plane using a vibrating sample magnetometer. It has a square loop with almost 100% remanance, indicating good perpendicular magnetic anisotropy in our films. This is required in our TR-MOKE measurement. The coercivity is ~1.97 kOe, which was optimized to meet the maximum applied field of our setup ability. The saturated magnetization ($M_s$) of the TbFe is ~420 emu cm$^{-3}$.

For comparison of through-plane thermal conductivity measurement using TR-MOKE and TDTR, one BP flake was also coated with an 80-nm Al transducer for TDRTR measurement using e-beam evaporation. The Al film serves as both a light absorber and a thermoreflectance transducer in the TDTR measurements[3,7-9]. Similar to the magnetic transducer deposition, a reference sample of 300-nm SiO$_2$ on silicon was also loaded inside the sputtering chamber together with the BP sample, for calibrating the thermal properties of the Al transducer. The electrical conductivity of the Al transducer was obtained from the 4-point probe technique. The thermal conductivity of the Al transducer was then determined with the Wiedemann-Franz Law. The heat capacity of the Al transducer was taken from literature[10]. For the TbFe alloy magnetic transducer, the thermal conductivity and heat capacity were obtained, respectively, from the in-plane beam offset and through-plane TR-MOKE measurements on the reference SiO$_2$ sample, which are detailed later in S2.4.

## S2.2 Beam spot size measurement in TR-MOKE and TDTR

All of the beam spot sizes used in this work were calibrated using the positive time delay beam offset measurements at high modulation frequency. Figure S4 presents a typical measured $V_{in}$ signal at 20 ps time delay at 9 MHz on $SiO_2$ reference samples coated with the 26.9-nm TbFe transducer layer. The solid lines are the best fit with a Gaussian function, used to extract $w_0$[3]. The fitted $w_0$ along two perpendicular directions are 3.59 and 3.62 $\mu$m, which gives an ellipticity for the spot of $1 \pm 0.01$ and thus excludes the contribution from the beam spots to the anisotropy of the 2D contour plots of the TR-MOKE signal.

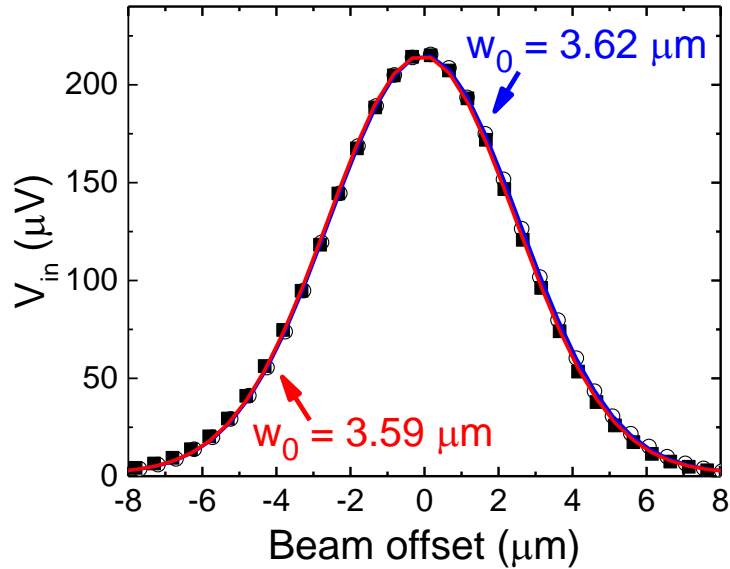

**Figure S4.** Beam spot size measurement on $SiO_2$ reference samples coated with the 26.9-nm TbFe transducer layer in TR-MOKE. The $V_{in}$ signals (open circles or filled squares) were taken with the 20× objective lens and $f = 9$ MHz at 20 ps along two orthogonal directions across the beam spot.

## S2.3 Sensitivity analysis of TR-MOKE signals

For the through-plane thermal conductivity measurement, the sensitivity of the ratio signal ($-V_{in} / V_{out}$) to a nominal parameter "$\sigma$" is defined as

$$S_z(\sigma) = \frac{\partial[\ln(-V_{in}/V_{out})]}{\partial[\ln(\sigma)]}, \tag{S1}$$

where $\sigma$ represents one of the geometrical parameters or thermal properties, and the subscript "z" denotes the through-plane direction. We analyze the full width at half maximum (FWHM) of the out-of-phase signal at negative time delay for the in-plane beam offset measurement; thus for the in-plane measurement, the sensitivity of the FWHM of $V_{\text{out}}$ is defined as

$$S_r(\sigma) = \frac{\partial[\ln(\text{FWHM})]}{\partial[\ln(\sigma)]} , \qquad (S2)$$

where the subscript "r" denotes the in-plane radial direction for general thermal properties. For BP crystals lacking of in-plane symmetry, $r$ refers to any in-plane direction and is equivalent to the notation of "a" or "c" when the direction of interest is along either the zigzag or armchair crystalline orientations of BP.

Figure S5 depicts the absolute sensitivity plots of through-plane TR-MOKE and in-plane beam offset measurements with both the TbFe and Al transducers. Based on the sensitivity analysis of the through-plane TR-MOKE measurements on reference SiO$_2$, at long time delay (> 100 ps), the heat capacity of the TbFe transducer ($C_\text{m}$) is the dominant parameter with the largest measurement sensitivity. Unlike $C_\text{m}$, the through-plane measurements are nearly insensitive to the thermal conductivity of TbFe magnetic transducer (the sensitivities to $\Lambda_{z,\text{m}}$ and $\Lambda_{r,\text{m}}$ are around $10^{-2}$ and $10^{-3}$, respectively). This allows the heat capacity of the TbFe transducer to be uniquely determined (Figure S5A). Figures S5B and S5C show the sensitivities of beam offset measurements of BP crystals coated with a 26.9-nm TbFe and with a 80-nm Al transducer layer, respectively, as a function of the sample in-plane thermal conductivity. The analysis is conducted for the case of a beam spot size of $w_0 = 3$ $\mu$m, a laser modulation frequency of $f = 1.6$ MHz, and at negative time delay of $-50$ ps. When the Al transducer is used for in-plane TDTR, the large in-plane thermal conductivity of Al ($\Lambda_{r,\text{m}} = 180$ W m$^{-1}$ K$^{-1}$) as compared to that of BP (30 W m$^{-1}$ K$^{-1}$ < $\Lambda_a$ or $\Lambda_c$ < 100 W m$^{-1}$ K$^{-1}$) causes significant heat spreading in the transducer layer, making the in-plane measurement less sensitive to $\Lambda_a$ or $\Lambda_c$ (Figure S4C). This situation is improved by in-plane TR-MOKE measurements. When the TbFe transducer is used, its lower thermal conductivity ($\leq 20$ W m$^{-1}$ K$^{-1}$) and smaller thickness will not only enhance the in-plane thermal transport in the BP crystal but will also make the measurement less sensitive to the transducer properties. Both of these effects are favored for reducing the uncertainty of the in-plane TR-MOKE measurements.

As shown in Figure S5B, the measurement sensitivity to $\Lambda_a$ or $\Lambda_c$ of BP, over the range from 30 to 110 W m$^{-1}$ K$^{-1}$, is one order of magnitude higher than that to $\Lambda_{r,m}$, and is three orders of magnitude higher than that to $\Lambda_{z,m}$ (not shown in Figure S5 due to the small amplitude of the measurement sensitivity to $\Lambda_{z,m}$, which is ~10$^{-4}$). Overall, the measurement is mostly sensitive to $\Lambda_{r,BP}$ and $C_{BP}$ among all the parameters, except for $w_0$, which we precisely characterized for each measurement to reduce the measurement uncertainty.

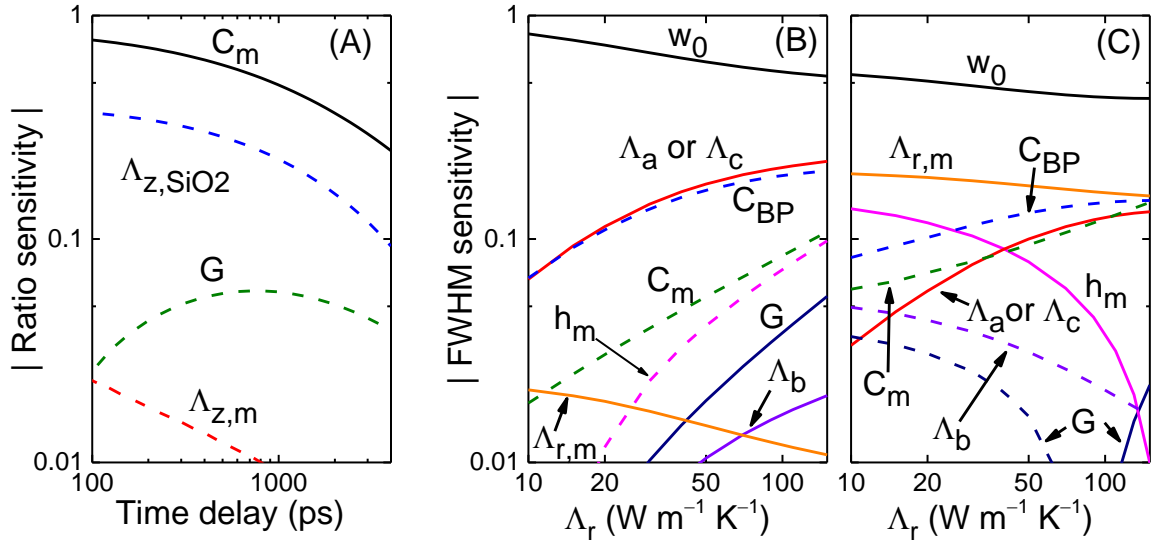

**Figure S5.** Sensitivity analysis of the through-plane and in-plane measurements to different parameters. (A) Absolute ratio sensitivities of through-plane TR-MOKE measurements to the thermal conductivity and heat capacity of a 26.9-nm TbFe transducer film on 300-nm SiO$_2$ reference. The measurement conditions are $w_0 = 12$ $\mu$m, and $f = 9$ MHz. Beam offset FWHM sensitivities of BP crystals coated with (B) the 26.9-nm TbFe magnetic transducer, and (C) the 81-nm Al metal transducer as a function of in-plane thermal conductivity $\Lambda_a$ or $\Lambda_c$ at measurement conditions of $w_0 = 3$ $\mu$m, and $f = 1.6$ MHz. For the sample substrate: $C_{SiO2} = 1.65$ J cm$^{-3}$ K$^{-1}$, $\Lambda_{z,SiO2} = 1.36$ W m$^{-1}$ K$^{-1}$ in (A); and $C_{BP} = 1.87$ J cm$^{-3}$ K$^{-1}$, $\Lambda_b = 5$ W m$^{-1}$ K$^{-1}$ in (B) and (C). For the metal transducer: $C_m = 3.1$ J cm$^{-3}$ K$^{-1}$, $\Lambda_{r,m} = 20$ W m$^{-1}$ K$^{-1}$ in (A) and (B), and $C_m = 2.42$ J cm$^{-3}$ K$^{-1}$; $\Lambda_{r,m} = 180$ W m$^{-1}$ K$^{-1}$ in (C). The time delay is set to –50 ps for in-plane beam offset measurements of BP crystals in both (B) and (C). Solid lines represent for sensitivities having positive values, while dashed lines represent negative sensitivities.

**S2.4 Control measurements of reference samples**

For thermal modeling analysis, some thermo-physical properties and geometric parameters need to be used as inputs. These include the heat capacity ($C_m$), thickness ($h_m$), and thermal conductivity (both in-plane $\Lambda_{r,m}$, and through-plane $\Lambda_{z,m}$) of the metallic transducer which can be either nonmagnetic Al or magnetic TbFe, and the through-plane thermal conductivity ($\Lambda_{z,BP}$, equivalent to $\Lambda_b$ for BP following the crystalline orientation notation as defined in Figure 1). To further reduce the uncertainties propagated from these input parameters and to validate the experimental approach, we carried out a series of control measurements using $SiO_2$ and silicon as reference samples.

In a typical TR-MOKE measurement, in-phase ($V_{in}$) and out-of-phase ($V_{out}$) signals are collected by an rf lock-in amplifier. Prior to each TR-MOKE measurement, the TbFe transducer is magnetized with an external magnet. The magnetization orientation of TbFe can be flipped by using different poles of the external magnet. Each sample is measured twice with the initial magnetization orientation of TbFe and its reverse, namely, $M+$ and $M-$. Both $V_{in}$ and $V_{out}$ change signs when the initial magnetization of TbFe is flipped from $M+$ to $M-$. Figure S6A shows the representative $V_{in}$ and $V_{out}$ signals of the $SiO_2$ reference sample for through-plane TR-MOKE measurements, taken with $M+$ (black open circles), and $M-$ (blue open circles). The difference of the two measurements is used for thermal analysis. For example, the corrected in-phase signal (red filled circles) is defined as $V_{in} = V_{in}^{M+} - V_{in}^{M-}$. This differential measurement approach effectively excludes any contribution from signals that are not associated with MOKE[8].

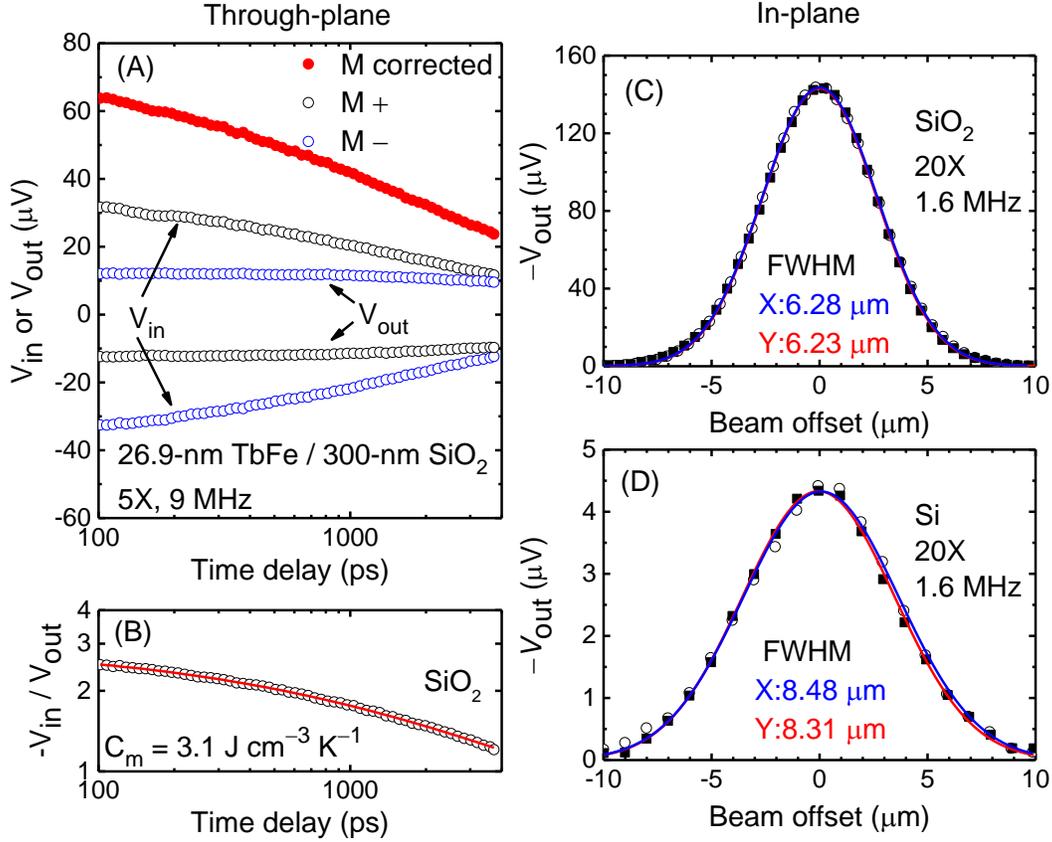

**Figure S6.** Validation measurements on reference samples of a 300-nm $SiO_2$ and silicon coated with the 26.9-nm TbFe magnetic transducer. (A) TR-MOKE $V_{in}$ and $V_{out}$ signals (open circles) on $SiO_2$ for measuring the through-plane thermal conductivity at 9 MHz and with $w_0 = 13$ μm. $M+$ (black) and $M-$ (blue) represent the opposite directions of the initial magnetization. (B) The ratio signal of $-V_{in}/V_{out}$ (open circles) and the best fit (solid red line) with the thermal model for the TbFe/$SiO_2$ reference sample[14]. (C) In-plane TR-MOKE measurement of the TbFe/$SiO_2$ reference sample. (D) In-plane TR-MOKE measurements of the TbFe/Si reference sample. The in-plane measurement signals taken along two orthogonal directions are nearly identical to each other, suggesting that the in-plane thermal transport is isotropic in both reference samples. Both the in-plane and through-plane thermal conductivities of $SiO_2$ are set as 1.36 W m$^{-1}$ K$^{-1}$ for (B) and (C). The fitted heat capacity and in-plane thermal conductivity of the TbFe transducer are $C_m = 3.1 \pm 0.2$ J cm$^{-3}$ K$^{-1}$ and $\Lambda_{r,m} = 17.4 \pm 1.4$ W m$^{-1}$ K$^{-1}$. The through-plane thermal conductivity of silicon in (D) is set as $\Lambda_{z,Si} = 110$ W m$^{-1}$ K$^{-1}$ as obtained from TDTR for fitting $\Lambda_{r,Si}$ of silicon, which is consistent with the best fit of $\Lambda_{r,Si} = 110 \pm 10$ W m$^{-1}$ K$^{-1}$ for the in-plane beam offset data of the silicon reference.

The procedures of signal processing and thermal analysis for $SiO_2$ and silicon are essentially the same as those used for BP flakes. Figure S6B plots the ratio data for the through-plane TR-MOKE measurement of the reference $SiO_2$ at 9 MHz with a 12-μm beam spot. A diffusive thermal model was used to fit for the heat capacity of the TbFe transducer[14]. The best fit gives $C_m = 3.1 \pm 0.2$ J cm$^{-3}$ K$^{-1}$ for the TbFe transducer and an interfacial thermal conductance of $G = 180 \pm 20$ MW m$^{-2}$ K$^{-1}$ between TbFe and SiO$_2$, which is close to the thermal conductance at Al/SiO$_2$ and CoPt/SiO$_2$ interfaces[1]. The thermal conductivity of the SiO$_2$ reference used in this fitting is $\Lambda_{SiO2} = 1.36 \pm 0.10$ W m$^{-1}$ K$^{-1}$, determined by measuring another SiO$_2$ reference diced from the same wafer with TDTR and the Al transducer. The thermal conductivity of the TbFe transducer can be obtained by fitting the through-plane TR-MOKE measurement of SiO$_2$ at short time delay from 30 ps to 700 ps, where the measurement sensitivity to $\Lambda_{z,m}$ is sufficient, see Figure S5A. The best fit gives $\Lambda_{z,m} = 16.2$ W m$^{-1}$ K$^{-1}$, which is typical for magnetic transducers that involve multi-layered structures. With the assumption of a 5% uncertainty in $w_0$, a 2% uncertainty in $C_{SiO2}$ and $C_{Si}$, an 8% uncertainty in $\Lambda_{SiO2}$ and $\Lambda_{Si}$, a 5% uncertainty in $h_m$ and a 1% uncertainty in $h_{SiO2}$, the total uncertainty of measured $C_m$ and $\Lambda_{z,m}$ are calculated to be 5.0% and 7.3%, respectively.

The in-plane thermal conductivity of the TbFe transducer was obtained based on the beam-offset TR-MOKE measurements of the SiO$_2$ reference with the 20× objective lens ($w_0 = 3.6$ μm) and a 1.6 MHz modulation frequency at −50 ps. Figure S6C illustrates the beam-offset TR-MOKE signals and best-fit curves taken along two orthogonal directions, which are nearly identical to each other and suggest that the in-plane thermal transport is isotropic in the SiO$_2$ reference. By modeling the FWHM of $V_{out}$ as a function of $\Lambda_{r,m}$ and fitting the measured FWHM, the in-plane thermal conductivity of the TbFe transducer is determined to be $\Lambda_{r,m} = 17.4 \pm 1.4$ W m$^{-1}$ K$^{-1}$. All the thermal properties of TbFe transducer layer (listed in Table S1) obtained from the control measurements of SiO$_2$ reference were treated as the known parameters for the analysis of TR-MOKE data on BP samples. The difference of the in-plane and through-plane thermal conductivities for TbFe is within the uncertainty of TR-MOKE measurements, suggesting that the TbFe transducer can be treated as thermally isotropic.

**Table S1.** Thermal properties of the TbFe magnetic transducer layer measured by TR-MOKE

| Thickness ($h_m$) | Heat capacity ($C_m$) | Through-plane $\Lambda_{z,m}$ | In-plane $\Lambda_{r,m}$ |
|---|---|---|---|
| [nm] | [J cm$^{-3}$ K$^{-1}$] | [W m$^{-1}$ K$^{-1}$] | [W m$^{-1}$ K$^{-1}$] |
| 26.9 | 3.1 ± 0.2 | 16.2 ± 1.2 | 17.4 ± 1.4 |

Before the 2D beam-offset measurement on BP flakes, we conducted control measurements on the thermally isotropic silicon reference sample (n-type, heavily doped, <100> surface), to further validate the method with high thermal conductivities. The in-plane TR-MOKE measurements of silicon were taken with a beam spot size of 3 $\mu$m and at 1.6 MHz, see Figure S6D. The measured in-plane thermal conductivities along two orthogonal directions are both $\Lambda_{r,Si} = 110 \pm 10$ W m$^{-1}$ K$^{-1}$ indicating no anisotropy. The through-plane thermal conductivity of the reference silicon, $\Lambda_{z,Si} = 110 \pm 9$ W m$^{-1}$ K$^{-1}$, was obtained from TDTR measurements ($f = 9$ MHz and $w_0 = 12$ $\mu$m) on another silicon reference sample cut from the same wafer and coated with an Al transducer. The measured value of the reference silicon sample is within the range of the thermal conductivity of heavily doped silicon wafers at room temperature reported in literature[11,12]. Excellent agreement between the in-plane thermal conductivities measured by beam-offset TR-MOKE and through-plane thermal conductivity measured by TDTR has been achieved, which demonstrates the experimental approach in this work is reliable and accurate for probing the in-plane thermal properties of high-thermal conductivity materials.

## S2.5 Steady-state heating

The temperature rise of steady-state heating ($\Delta T_{ss}$) in TR-MOKE measurement can be estimated using[14]

$$\Delta T_{ss} < \frac{\alpha_m P}{2\pi w_0 \Lambda} \ , \tag{S3}$$

where $\alpha_m$ is the absorption coefficient of the metal transducer film, $\Lambda$ is the thermal conductivity of the substrate, $w_0$ is the beam spot radius, and $P$ is the total power of the laser beam. Since the absorption coefficient of the TbFe film is unknown, we estimate its value by comparing the reflectivity of the probe beam from the BP samples coated with TbFe and from the reference SiO$_2$ coated with Al. When the same power of the probe beam is used for both cases, the direct current (dc) voltage ($V_{dc}$) of the reflected probe signal from Al and TbFe surfaces are $V_{dc,Al} = 140$ mV and $V_{dc,TbFe} = 87$ mV, respectively. This allows $\alpha_{TbFe}$ to be derived based on

$$\frac{V_{dc,TbFe}}{V_{dc,Al}} = \frac{1-\alpha_{TbFe}}{1-\alpha_{Al}}, \tag{S4}$$

where $\alpha_{Al} = 0.1$ for an optically opaque Al film. $\alpha_{TbFe}$ is thus calculated to be ≈0.44. The geometric mean of the 3D anisotropic thermal conductivity of BP is used to estimate $\Delta T_{ss}$, which gives $\Lambda_{mean} = \sqrt[3]{\Lambda_a \times \Lambda_b \times \Lambda_c} = 23$ W m$^{-1}$ K$^{-1}$. With a pump power of 12 mW and a probe power of 6 mW, $\Delta T_{ss}$ is calculated to be 10 K for BP using Eq. (S3), which validates the use of the linear heat transfer model[14].

## S2.6 Laser drift correction by monitoring the beam power

We minimize the artifacts due to the laser power and wavelength drift by correcting the TR-MOKE signal with the power signals of both the pump and probe beams monitored with a fast-response power meter and a photodiode detector, throughout the entire scanning measurement duration, respectively. Figure S7 shows a typical $V_{dc}$ contour of the probe beam obtained by a photodiode detector during TR-MOKE beam offset 2D scanning on sample BP-1. The deviation of $V_{dc}$ induced by the drift of the laser is within 10%. All final $V_{out}$ FWHM signals have been corrected with $V_{dc}$ by scaling the $V_{out}$ contour signal with $V_{dc}$.

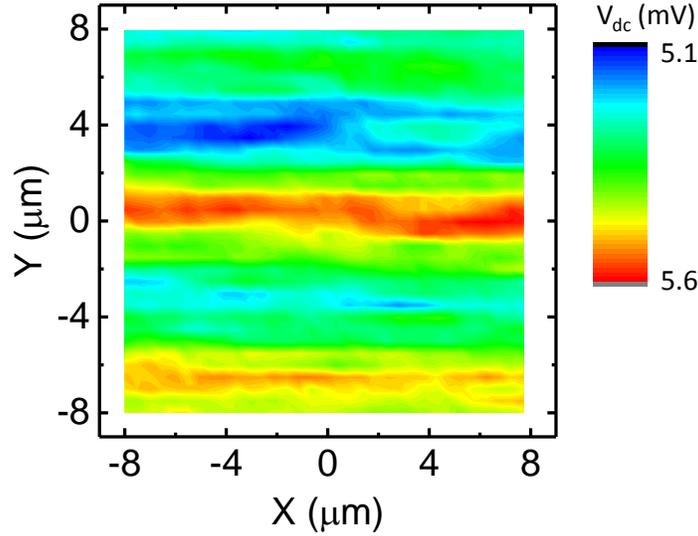

**Figure S7.** $V_{dc}$ contour of probe beam obtained by a photodiode detector during TR-MOKE beam offset 2D scanning measurements on the BP-1 sample. This $V_{dc}$ signal is used to correct the laser drift artifact.

# S3. First-principles calculation for phonon transport

## S3.1 Solutions of Peierls-Boltzmann transport equation

We performed first-principles-based Peierls-Boltzmann transport equation calculations to predict the thermal conductivity of both bulk and monolayer BP.

To calculate the thermal conductivity along $\alpha$ ($= a, b, c$) direction, a small temperature gradient was applied along $\alpha$ direction, $\frac{\partial T}{\partial x^\alpha}$, to perturb the phonon population from the equilibrium Bose-Einstein distribution, $n_{qs}^0$ to the non-equilibrium one $n_{qs}$, where $qs$ stands for the $s$-th phonon mode at $q$ in the first Brillouin zone. When the steady state was achieved, the heat flux can be expressed as the summation of the contributions from all phonon modes through

$$J^\alpha = \frac{1}{(2\pi)^3}\sum_s \int \hbar \omega_{qs} v_{qs}^\alpha n_{qs} d\boldsymbol{q}, \qquad (S5)$$

where $\hbar$ is the Planck constant, $\omega_{qs}$ and $v_{qs}$ are the frequency and group velocity of phonon mode $qs$. The phonon frequency $\omega_{qs}$ and $\boldsymbol{v}_{qs} = \nabla \omega_{qs}$ obtained from the phonon dispersion of the crystal, which is related to the second-order harmonic force constants of the crystal, $\phi$. After $J^\alpha$ was calculated with the contributions from all phonon modes, the macroscopic thermal conductivity was then derived using the Fourier's law of heat conduction, $\Lambda_{\alpha\alpha} = J^\alpha/(\frac{\partial T}{\partial \alpha})$. In order to evaluate the heat flux driven by the applied small temperature gradient, non-equilibrium phonon distribution of each phonon mode in Eq. (S5) needs to be determined.

To determine the non-equilibrium phonon population distribution $n_{qs}$ in Eq. (S5), it is convenient to express $n_{qs}$ in the form of $n_{qs}^0 + n_{qs}^0(n_{qs}^0 + 1)F_{qs}^\alpha \frac{\partial T}{\partial x^\alpha}$ with deviation function $F_{qs}^\alpha$ [15,16]. The linearized PBTE, which describes the balance between phonon diffusion driven by the applied temperature gradient and phonon scattering, was then solved for the non-equilibrium phonon distribution function, or its equivalent $F_{qs}$. In this work, we considered the three-phonon scattering, and the corresponding PBTE can be expressed as[15,16]:

$$v_{qs}^\alpha \frac{\partial n_{qs}^0}{\partial T} = \sum_{q's',q''s''}\left[W_{qs,q's'}^{q''s''}\left(F_{q''s''}^\alpha - F_{q's'}^\alpha - F_{qs}^\alpha\right) + \frac{1}{2}W_{qs}^{q's',q''s''}\left(F_{q''s''}^\alpha + F_{q's'}^\alpha - F_{qs}^\alpha\right)\right],$$

(S6)

where $W_{qs,q's'}^{q''s''}$ and $W_{qs}^{q's',q''s''}$ are the equilibrium transition probabilities for three-phonon annihilation and decay scattering processes, respectively, which are expressed as

$$W_{qs,q's'}^{q''s''} = 2\pi n_{qs}^0 n_{q's'}^0 \left(n_{q''s''}^0 + 1\right) |V_3(-qs, -q's', q''s'')|^2$$

$$\times \delta(\omega_{qs} + \omega_{q's'} - \omega_{q''s''})\Delta(q + q' - q'' + G) \tag{S7}$$

$$W_{qs}^{q's',q''s''} = 2\pi n_{qs}^0 \left(n_{q's'}^0 + 1\right)\left(n_{q''s''}^0 + 1\right) |V_3(-qs, q's', q''s'')|^2$$

$$\times \delta(\omega_{qs} - \omega_{q's'} - \omega_{q''s''})\Delta(q - q' - q'' + G) \tag{S8}$$

Here, the $\delta$ function and the $\Delta$ function denote the energy conservation condition $\omega_{qs} \pm \omega_{q's'} - \omega_{q''s''} = 0$ and the momentum conservation condition $q \pm q' - q'' + G = 0$ for the three-phonon scattering process. The + and − signs represent the annihilation and decay processes, respectively. The $G$ vector represents a reciprocal vector. The scattering is normal process when $G = 0$, while it is Umklapp process when $G \neq 0$. The three-phonon scattering matrix $V_3$ is written as

$$V_3(qs, q's', q''s'') = \left(\frac{\hbar}{8N_0\omega_{qs}\omega_{q's'}\omega_{q''s''}}\right)^{\frac{1}{2}} \sum_\tau \sum_{R'\tau'} \sum_{R''\tau''} \sum_{\alpha\beta\gamma} \psi_{0\tau,R'\tau',R''\tau''}^{\alpha\beta\gamma}$$

$$\times \exp(iq' \cdot R' + iq'' \cdot R'') \frac{\varepsilon_{qs}^{\tau\alpha} \varepsilon_{q's'}^{\tau'\beta} \varepsilon_{q''s''}^{\tau''\gamma}}{\sqrt{M_\tau M_{\tau'} M_{\tau''}}}, \tag{S9}$$

where $\varepsilon$ is the eigenvector of the dynamical matrix, $M$ is the atom mass, $N_0$ is the number of unit cells, and the $(R, \tau)$ represents the $\tau$-th atom in the unit cell located in $R$.

In the calculation for bulk (monolayer) BP, we sampled the first Brillouin zone using a $N \times N \times N$ ($N \times N \times 1$) q-mesh. For each phonon mode in the discrete point **q**, we had to find out all possible phonon pairs to scatter with it, which makes both the momentum conservation and energy conservation held after scattering with the given phonon mode. Here, we employed the phase-space method[15,17]. In this method, the first Brillouin zone of bulk (monolayer) BP is discreted to $N \times N$ ($N \times 1$) one-dimensional sub-space, and for the given phonon mode, the possible phonon pairs that can interact with it are searched along the one-dimensional space.

The PBTE, Eq. (S6), clearly shows that the phonon distribution function of each phonon mode is coupled with the distribution of other phonon modes, which indeed makes the PBTE challenging to solve. One way to solve the PBTE is to use the so-called single-mode relaxation time approximation (SMRTA) by assuming each phonon mode is decoupled with other modes, or $F_{q's'}^\alpha = F_{q''s''}^\alpha = 0$. Then, the thermal conductivity is written as

$$\Lambda_{\alpha\alpha} = \frac{\hbar^2}{N_0 \Omega k_B T^2} \sum_{qs} \omega_{qs}^2 (v_{qs}^\alpha)^2 n_{qs}^0 (n_{qs}^0 + 1)\tau_{qs}, \qquad (S10)$$

where $\Omega$ is the volume of the primitive unit cell, $\tau_{qs}$ is the phonon relaxation time from the SMRTA, which is given by

$$\tau_{qs} = \frac{n_{qs}^0(n_{qs}^0+1)}{\sum_{q's',q''s''}\left[W_{qs,q's'}^{q''s''} + \frac{1}{2}W_{qs}^{q's',q''s''}\right]}. \qquad (S11)$$

Due to the simplicity of SMRTA, it has been widely used in literature to analyze the thermal conductivity data. However, SMRTA cannot distinguish the resistive Umklapp process and the normal process, which does not directly provide the resistance to the heat flow, leading to underestimation of the thermal conductivity[18].

With recent advances, the set of linear equations Eq. (S6), with respect to $F_{qs}^\alpha$, can be self-consistently solved through iterative method[15]. Here we employed the biconjugate gradient stabilized method (Bi-CGSTAB)[19], a variant of the conjugate gradient algorithm, to iteratively solve it. After $F_{qs}^\alpha$ was calculated, the thermal conductivity can be expressed as

$$\Lambda_{\alpha\alpha} = \frac{1}{N_0 \Omega} \sum_{qs} \hbar\omega_{qs} v_{qs}^\alpha n_{qs}^0 (n_{qs}^0 + 1) F_{qs}^\alpha, \qquad (S12)$$

By strictly solving Eq. (S6) iteratively, the coupling among different phonon modes has been naturally taken into account.

To quantify the importance that the temperature gradient collective scattering behavior of all the phonon modes plays on anisotropic phonon transport, we rewrite Eq. (S12) into the form of Eq. (S10) to defining the effective phonon relaxation time $\tilde{\tau}_{qs}^\alpha$ in the iterative approach, which is expressed as

$$\tilde{\tau}_{qs}^\alpha = (k_B T^2/\hbar) \cdot (F_{qs}^\alpha / v_{qs}^\alpha \omega_{qs}). \qquad (S13)$$

## S3.2 Extracting interatomic force constants from first-principles calculations

To obtain phonon dispersion and the phonon-phonon scattering rates in Eq. (S7) and Eq. (S8) for computing the thermal conductivity, we need to calculate both the second-order harmonic force constants and the third-order anharmonic force constants of bulk and monolayer BP. We applied the first-principles calculations to obtain their equilibrium crystal structures and then to extract the interatomic force constants corresponding to the equilibrium crystal structures.

Our first-principles calculations were carried out using the Vienna *ab initio* Simulation Package (VASP)[20]. The projector augmented wave pseudopotential with PBE functional[21] is employed. For bulk BP, we took into account the van der Waals interaction by using the dispersion correction (DFT-D) proposed by Grimme[22]. We used periodic boundary conditions throughout the study. In the simulation of monolayer BP, a 15 Å thick vacuum region is inserted between two BP layers in neighboring images. Figure S8 illustrates the unit cells for bulk and monolayer BP. The kinetic-energy cut-off for the plane-wave basis set is set to be 500 eV and a $12\times12\times6$ ($12\times12\times1$) Monkhorst-Pack mesh is used to sample the reciprocal space of bulk (monolayer) BP. When we further refine these two parameters, the energy change is smaller than 1 meV per atom. The crystal structure, including the lattice constants and atom coordinates, is relaxed through the conjugate gradient algorithm until the stress within the material is zero and the atomic forces are smaller than $1\times10^{-5}$ eV Å$^{-1}$. The optimized lattice constants, which are denoted as *a*, *b* and *c* for the zigzag, through-plane and armchair directions, respectively, are summarized in Table S2.

**Table S2**. Lattice constants of both bulk and monolayer BP

|  | Method | *a* (Å) | *b* (Å) | *c* (Å) |
|---|---|---|---|---|
|  | DFT-D (this work) | 3.321 | 10.477 | 4.427 |
| Bulk | DFT-D (Ref. [23]) | 3.30 | 10.43 | 4.40 |
|  | Exp. (Ref. [24]) | 3.3133 | 10.473 | 4.374 |
| Monolayer | DFT-D (this work) | 3.306 | - | 4.572 |

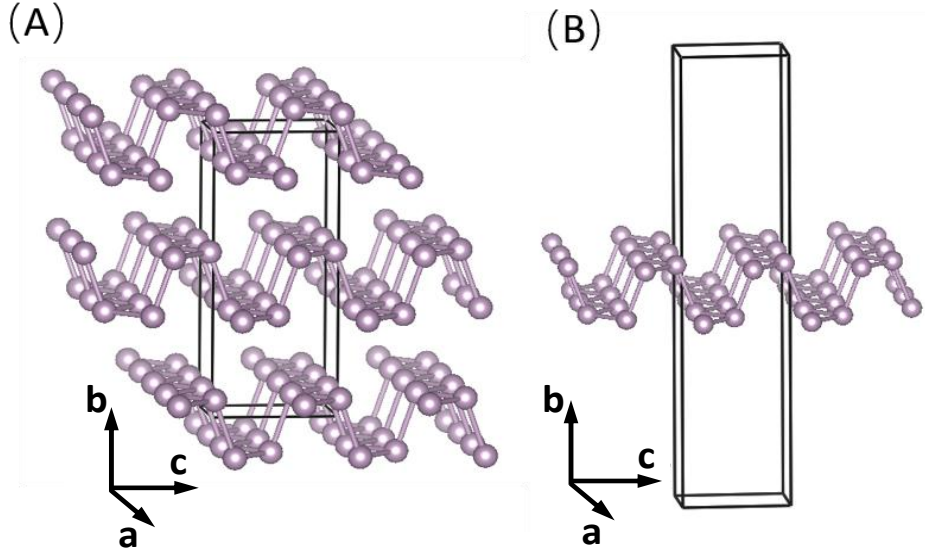

**Figure S8**. Crystal structure of (A) bulk and (B) monolayer BP.

With the obtained equilibrium crystal structures, we extracted the interatomic force constants using the supercell-based method in the first-principles calculations[25]. The force component in $\alpha$ direction on each atom in a supercell is expressed as[25]

$$F_{R\tau}^{\alpha} = -\sum_{R'\tau'\beta} \phi_{R\tau,R'\tau'}^{\alpha\beta} u_{R'\tau'}^{\beta} - \frac{1}{2}\sum_{R'\tau'\beta}\sum_{R''\tau''\gamma} \psi_{R\tau,R'\tau',R''\tau''}^{\alpha\beta\gamma} u_{R'\tau'}^{\beta} u_{R''\tau''}^{\gamma} + \cdots \qquad (S14)$$

where $u$ is displacement of an atom away from its equilibrium position. When calculating the second-order harmonic force constants, we displaced one atom in a supercell by a small displacement $\Delta u = 0.02$ Å away from its equilibrium position along $\pm x$, $\pm y$ and $\pm z$ directions, and then recorded the forces of all atoms in the supercell. With the recorded forces, we extracted the second-order harmonic force constants by fitting the displacement-force data set according to Eq. (S14).

The cutoff of the harmonic interactions is chosen to be 3.0 $a$ (~ 10 Å). For monolayer BP, we used the supercell made up of 6×6×1 primitive unit cells to generate displacement-force data set. For bulk phosphorus, in order to take the interlayer interaction into account, more than one layers should be included in the supercell, leading to the computational challenges due to the large amount of atoms in the simulation. To avoid employing big supercells to obtain the displacement-force data corresponding to the long-range interlayer interaction, we used two kinds of small supercells with different dimensions (6×6×1 and 4×4×2 conventional unit cells) to generate two sets of displacement-force data sets, which are fitted simultaneously to extract the second-order harmonic force constants.

With the extracted harmonic force constants, the phonon dispersion of BP is calculated. Figure S9 shows that the calculated phonon dispersion of bulk BP agrees reasonably well with the experimental data from the inelastic neutron scattering measurements[26].

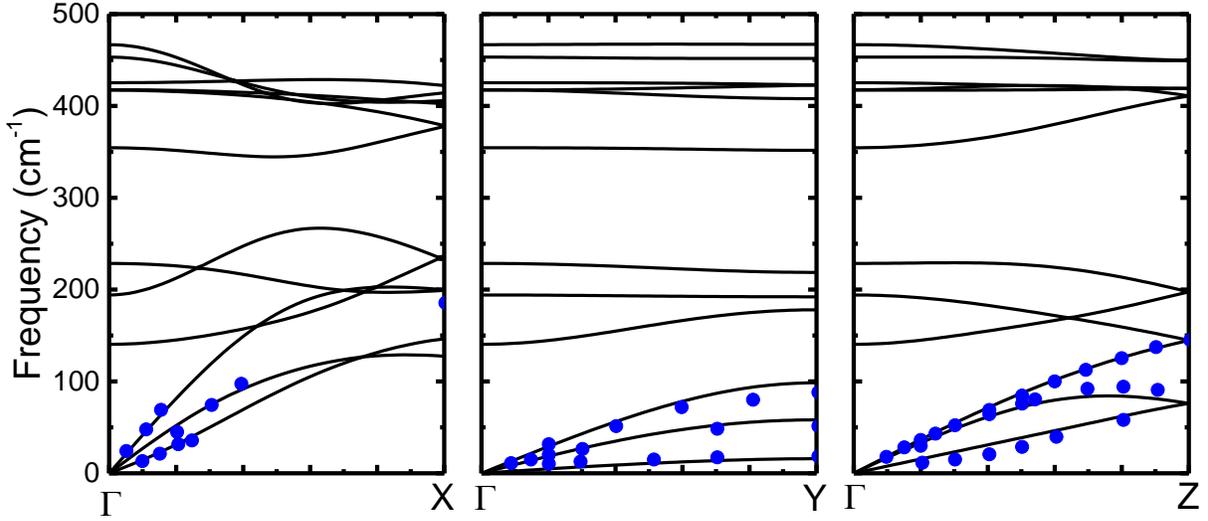

**Figure S9**. The phonon dispersion of bulk BP. Black lines are the calculated phonon dispersion using the second-order harmonic force constants extracted from first-principles calculations. Blue dots are the experimental results from inelastic neutron scattering measurement[26].

Similarly, to extract the third-order anharmonic force constants for bulk (monolayer) BP, we displaced two atoms in a 4×3×2 (4×3×1) conventional unit cell with a distance of 0.02 Å along different directions simultaneously. With the recorded force information on all atoms in the supercell, the third-order force constants was then calculated using the finite-difference scheme[27]:

$$\psi^{\alpha\beta\gamma}_{R\tau,R'\tau',R''\tau''} = \frac{1}{4\Delta u^2}\Big[-F^{\alpha}_{R\tau}\Big(u^{\beta}_{R'\tau'} = \Delta u, u^{\gamma}_{R''\tau''} = \Delta u\Big) + F^{\alpha}_{R\tau}\Big(u^{\beta}_{R'\tau'} = \Delta u, u^{\gamma}_{R''\tau''} = -\Delta u\Big) +$$
$$F^{\alpha}_{R\tau}\Big(u^{\beta}_{R'\tau'} = -\Delta u, u^{\gamma}_{R''\tau''} = \Delta u\Big) - F^{\alpha}_{R\tau}\Big(u^{\beta}_{R'\tau'} = -\Delta u, u^{\gamma}_{R''\tau''} = -\Delta u\Big)\Big]. \quad (S15)$$

As the calculated thermal conductivity is highly sensitive to the cutoff used for the calculation of the third-order force constants, we carefully tested the choice of cutoff. Figure S10 shows the calculate thermal conductivity of bulk BP with different cutoff using a 13×13×13 q-point phonon mesh on the first Brillouin zone. It is clearly seen that a too small cutoff could substantially overestimate the thermal conductivity along all directions. When the cutoff is larger than 5.55 Å, the thermal conductivity variation is smaller than 10%. Therefore, we chose 5.55 Å as the cutoff for anharmonic third-order force constants for further calculations.

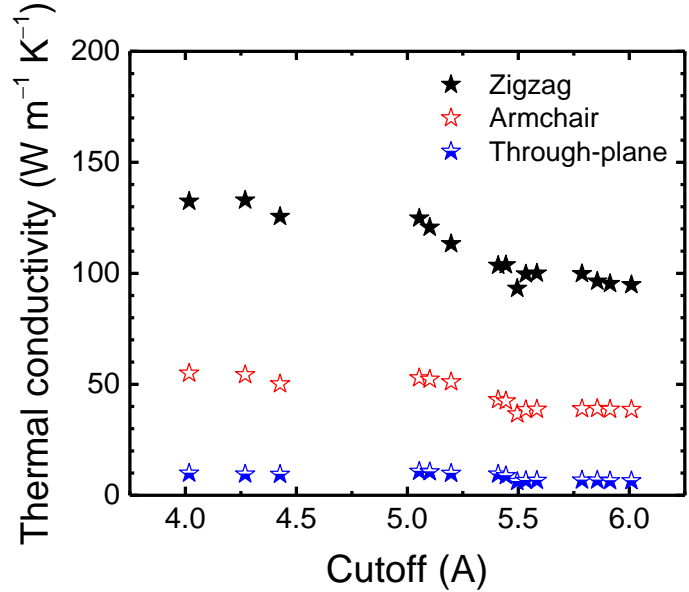

**Figure S10**. The calculated thermal conductivity of bulk BP as a function of the cutoff used for the calculation of third-order anharmonic force constants.

### S3.3 Calculation of thermal conductivity in bulk and monolayer black phosphorus

Before calculating the temperature-dependent thermal conductivity of monolayer and bulk BP, we carefully examined how the thermal conductivity changes with the q-point mesh, which is used to evaluate the integrals in Eq. (S5) and Eq. (S6), at room temperature. We tested a series of $N \times N \times N$ meshes and found that the thermal conductivity is almost constant (within 1%) when $N > 41$ for monolayer one and $N > 15$ for bulk one, as shown in Figure S11. Therefore, we used $N = 61$ and $N = 19$ in the remaining thermal conductivity calculations for monolayer and bulk BP, respectively.

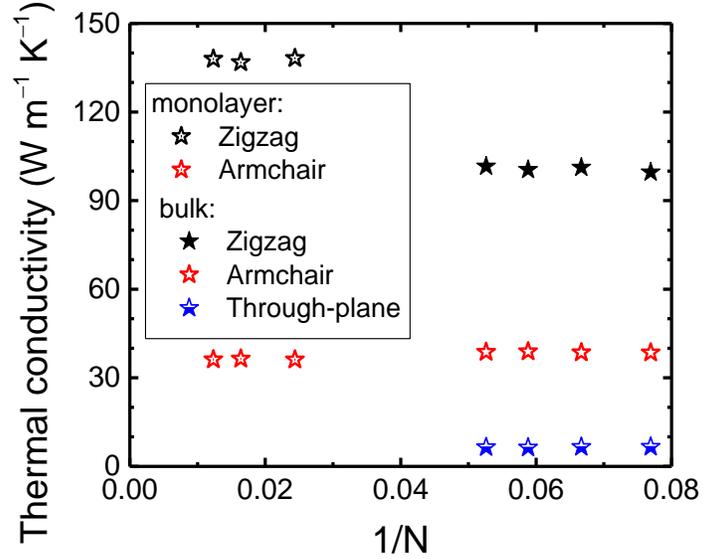

**Figure S11**. The calculated thermal conductivity of BP as a function of the ($N×N×N$) q-point mesh. We test the meshes with $13×13×13$, $15×15×15$, $17×17×17$, $19×19×19$ points for bulk BP and $41×41×41$, $61×61×61$, $81×81×81$ points for monolayer one.

The computational results for the thermal conductivity of monolayer BP have been reported in literature. Table S3 summarizes our results in comparison with the data reported in literature. Since some previous works employed single-mode relaxation approximation (RTA) to calculate the thermal conductivity, we also list our SMRTA results in Table S3. In general, our results are close to Zhu et al[28] as well as Jain and Mcgaughey[29].

**Table S3**. Summary of the calculated thermal conductivity of monolayer BP

| Reference | Method | $\Lambda_{zigzag}$ | $\Lambda_{armchair}$ |
| --- | --- | --- | --- |
| Zhu (Ref. [28]) | RTA | 95 | 24 |
| Jain (Ref. [29]) | RTA | 84 | 30 |
| Qin (Ref. [30]) | RTA | 30 | 14 |
| This work | RTA | 94 | 29 |
| Jain (Ref. [29]) | Iterative | 110 | 36 |
| This work | Iterative | 137 | 36 |

With the iterative approach to solve PBTE, our calculated thermal conductivity of bulk BP are 102, 6.5, and 39 W m$^{-1}$ K$^{-1}$ along the zigzag, interlayer (through-plane), and armchair directions, respectively. These values compare very favorably with the measured thermal conductivity. Figure S12 presents the temperature-dependent thermal conductivity. We noticed that a recent study[31] reported their calculated thermal conductivity to be about 155, 45, and 65 W m$^{-1}$ K$^{-1}$ for the zigzag, through-plane and armchair directions, which are substantially higher than our computational results and measured values. This is probably due to a too small cutoff for the third-order force constants used in their study, which was set to be only the fourth nearest neighbors (~3.3 Å). With such a small cutoff, the interlayer interaction is totally ignored, leading to an unphysically high value for the through-plane thermal conductivity.

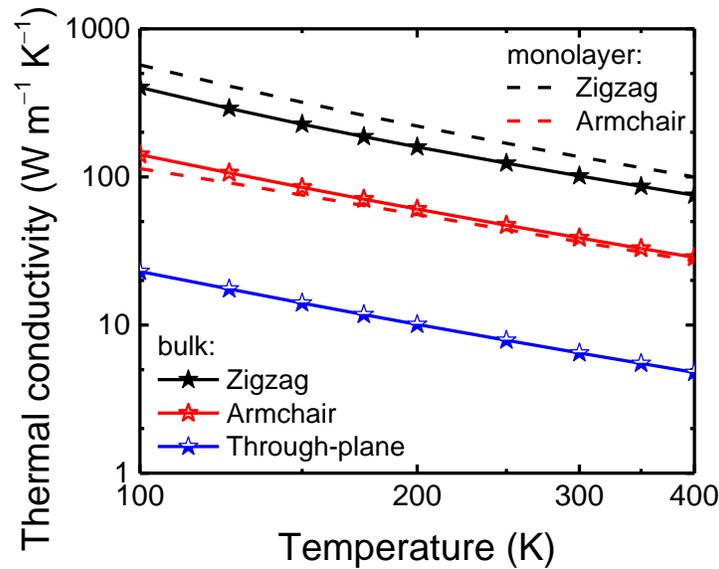

**Figure S12**. The calculated thermal conductivity of both bulk and monolayer BP as a function of temperature.